\documentclass{ws-ijam}

\def\thebibliography#1{
 \list
 {[\arabic{enumi}]}{\settowidth\labelwidth{[#1]}\leftmargin\labelwidth
 \advance\leftmargin\labelsep
 \usecounter{enumi}}
 \def\newblock{\hskip .11em plus .33em minus -.07em}
 \sloppy
 \sfcode`\.=1000\relax}
 
\usepackage[square]{wsnatbib}   

\usepackage{graphicx}
\usepackage{empheq}
\usepackage{amsmath,amssymb}
\usepackage{wasysym}
\usepackage{color}
\usepackage{siunitx} 
\usepackage{multirow} 

\begin{document}

\markboth{N. M. Sangtani Lakhwani et al}
{Lattice Boltzmann Method simulations of high Reynolds number flows past porous obstacles}

\catchline{}{}{}{}{}

\title{Lattice Boltzmann Method simulations of high Reynolds number flows past porous obstacles}

\author{N. M. Sangtani Lakhwani}
\address{SFMG, Department of Mechanical Engineering, The University of Sheffield, Sheffield, UK}

\author{F.C.G.A Nicolleau\footnote{
Corresponding author.}} 
\address{SFMG, The University of Sheffield, Sheffield, UK}

\author{W. Brevis}
\address{Department of Hydraulics and Environmental Engineering \& Mining Engineering, Pontifical Catholic University of Chile, Santiago, Chile}

\maketitle

\begin{history}
\received{date}
\accepted{date}
\end{history}

\date{\today}

\begin{abstract}
Lattice Boltzmann Method (LBM) simulations for turbulent flows over a fractal and non-fractal obstacles are presented.  
The wake hydrodynamics are compared and discussed in terms of flow relaxation, Strouhal numbers and wake length for different Reynolds numbers.
Three obstacle topologies are studied, Solid (SS), Porous Regular (PR) and Porous Fractal (FR). 
In particular we observe that
the oscillation present in the case of the solid square can be
annihilated or only pushed downstream depending on the topology ot the porous obstacle.

The Lattice Boltzmann Method (LBM) is implemented over a range of four Reynolds numbers from 12352 to 49410. 
The suitability of LBM for these high Reynolds number cases is studied. Its results are compared to available experimental data and published literature.
Compelling agreements between all three tested obstacles show a significant validation of LBM as a tool to investigate high Reynolds number flows in complex geometries. This is particularly important as the LBM method is much less time consuming than a classical Navier-Stokes equation based computing method and high Reynolds numbers need to be achieved with enough details (i.e. resolution) to predict for example canopy flows.

\end{abstract}


\maketitle

\section{Introduction}
\label{secintro}

With ever growing levels of urbanisation across the globe, a good understanding of complex flows (particularly flows past porous obstacles) is paramount to reduce pollution in major cities and prevent unwanted aerodynamic loading on structures. The multi-scale nature of not only urban construction but that of natural environments requires a more complex modelling system be employed. Fractal geometries 
have only recently been investigated in turbulent flows, \citep{Coleman-Vassilicos-2008,Laizet-Vassilicos-2012,Wei-et-al-2016-FDR,Nicolleau-et-al-JOT-2011}
their multi-scale properties make them the logical choice for parametric studies for modelling and simulating flows involving such complex geometries. More precisely we should be talking about pre-fractals as they involve only few iterations of a fractal pattern. In this study our obstacle will be based on Sierpi\'nski's carpet with a maximum of 3 scales.

Additionally, in recent years the usage of Lattice Boltzmann Methods (LBM) for Computational Fluid Dynamics (CFD) has increased, since LBM offers better computational efficiency and speed over Navier-Stokes equation based CFD. However,  LBMs still need to be benchmarked since they work differently than  the `classical' methods in particular macroscopic quantities of the flow are extracted using a probabilistic model of the flow at microscopic scales. 
Flows through porous obstacles which offer a complex range of turbulence scales interaction are challenging candidates for such validations. We propose to validate LBM results with experimental data from flows passing through three obstacle topologies: Solid (SS), Porous Regular (PR) and Porous Fractal (FR).
\\[2ex]
The paper is organised as follows: in \S~\ref{secmethod} we describe the simulation domain set up, its notations and its parameters. 
The results are then presented and discussed in \S~\ref{secresults}.
The main conclusions are summarised in  \S~\ref{seconcl}.

\section{Simulation set up}
\label{secmethod}

\begin{figure}[h]
	\centering
	\includegraphics[width=\textwidth]{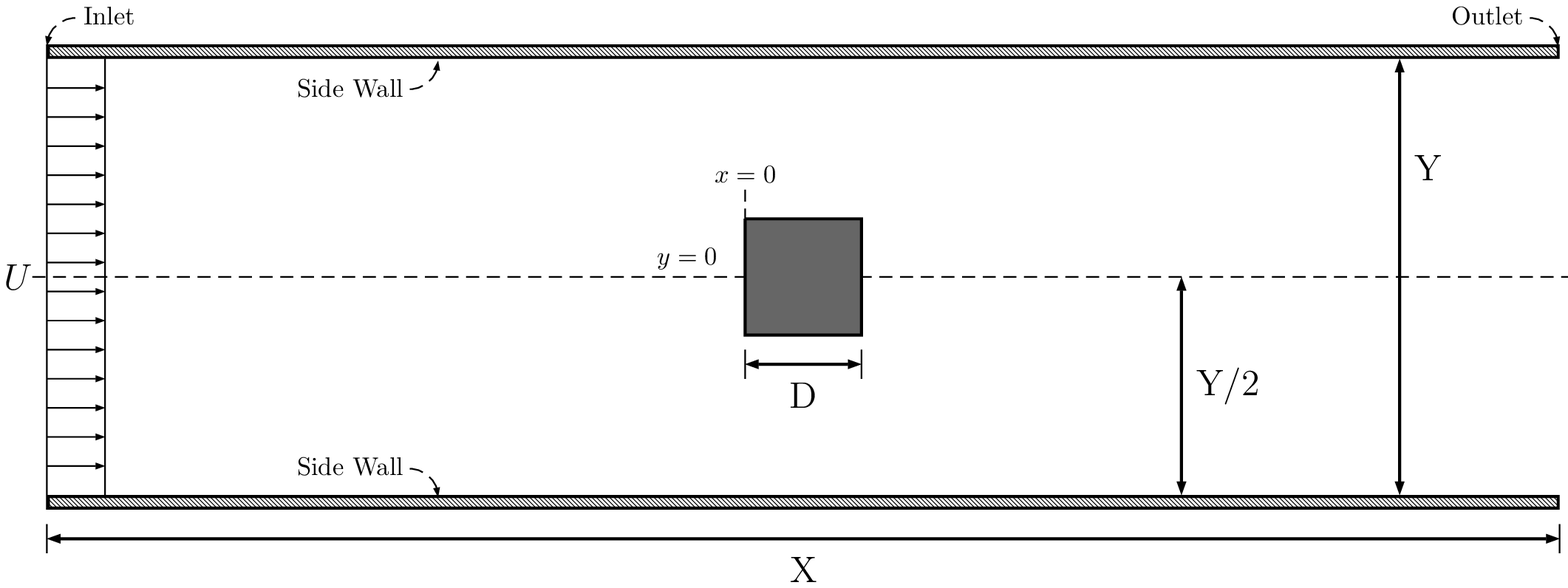}	
	\includegraphics[width=\textwidth]{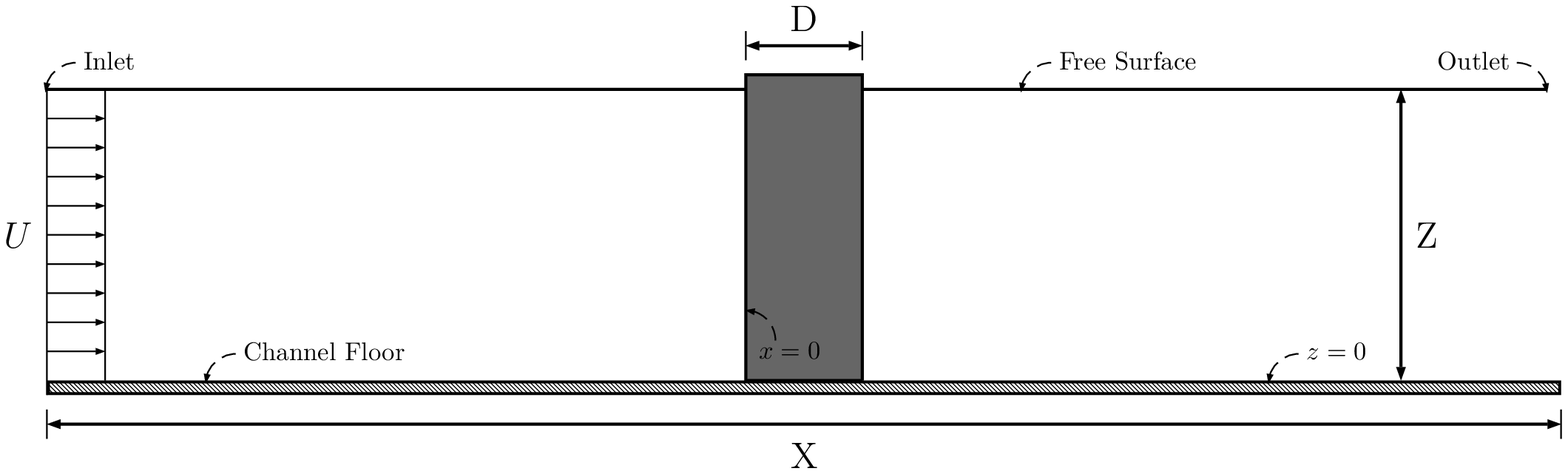}
	
	\caption{Schematic of the flume and geometry of the numerical domain}
	\label{FlumeFig}
\end{figure}

\begin{figure}[h]
	\centering
	\includegraphics[width=0.32\textwidth]{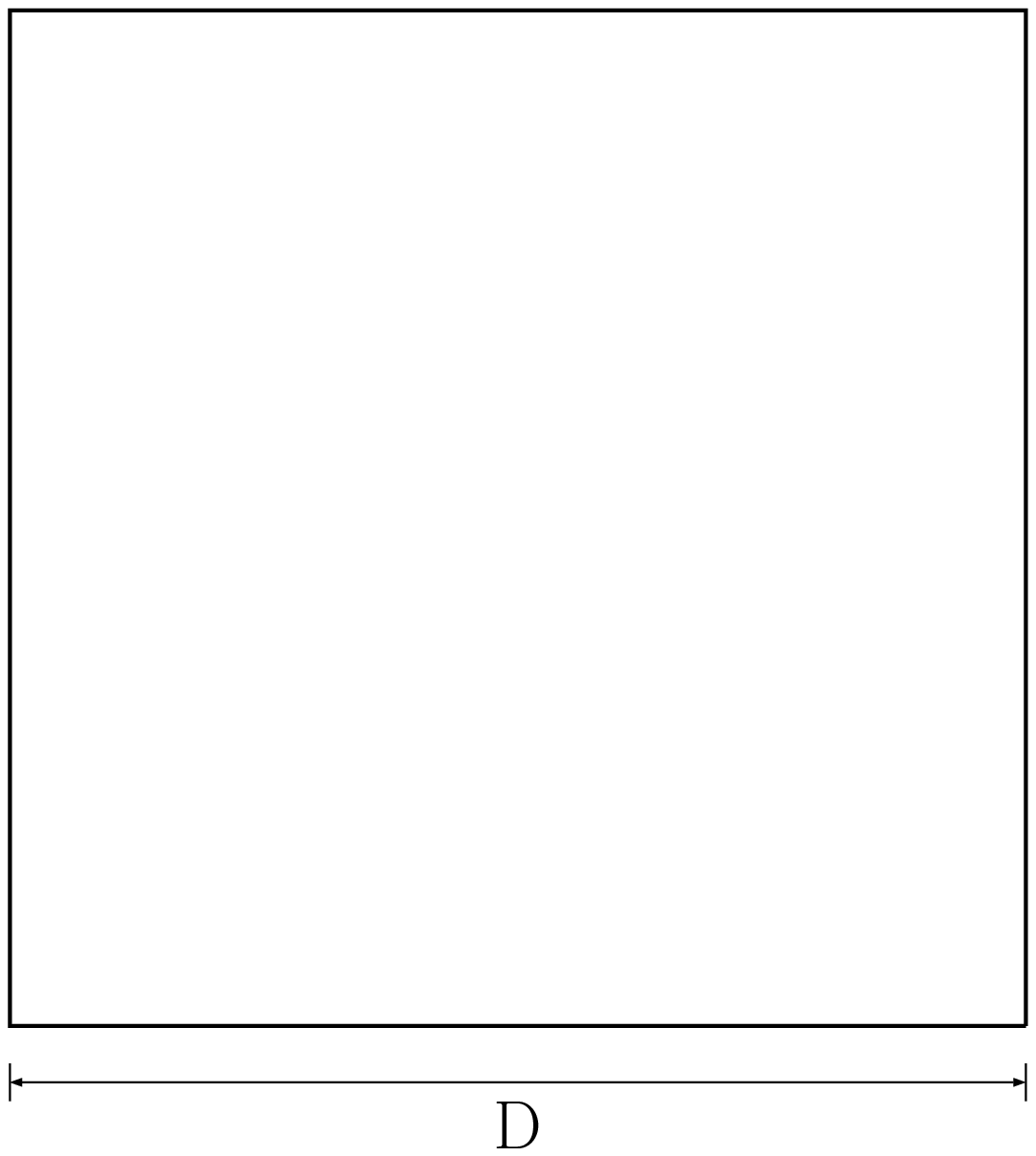}
	\includegraphics[width=0.32\textwidth]{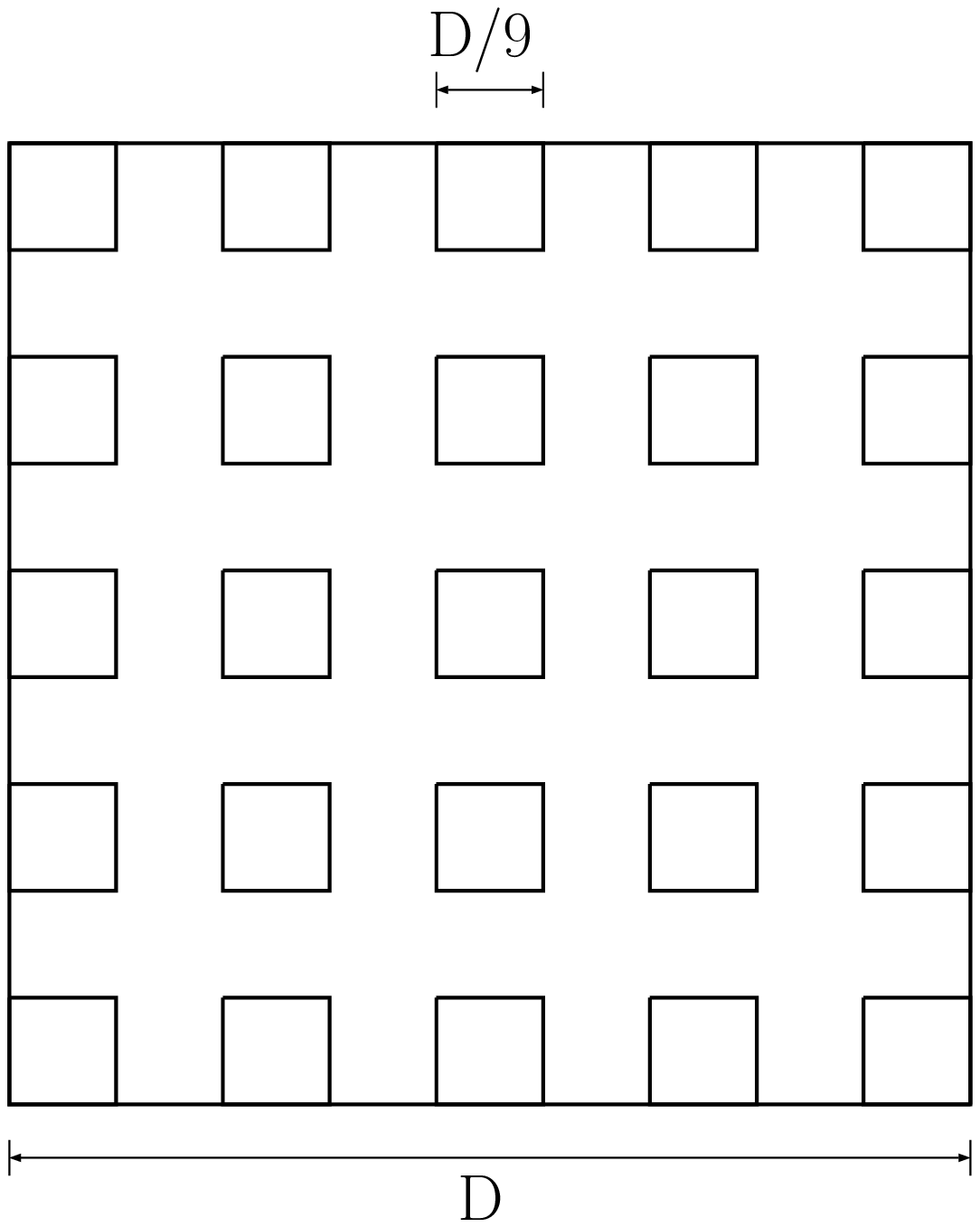}
	\includegraphics[width=0.32\textwidth]{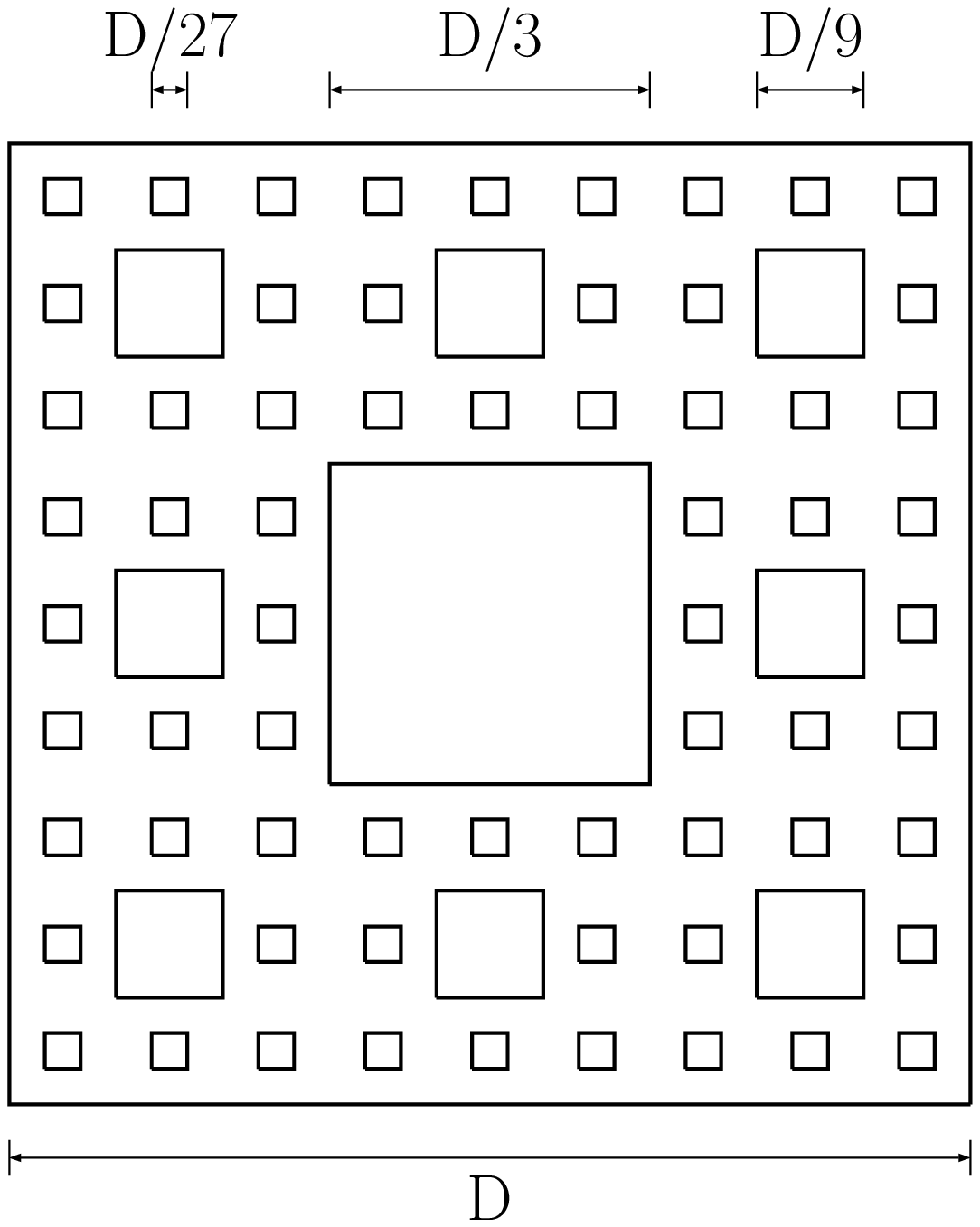}

	\caption{Top view of the obstacles being used for this investigation and their geometries. (Left) Solid Square, (Middle) Porous Regular, (Right) Porous Fractal}
	\label{ObstaclesFig}
\end{figure}

As the main purpose of this investigation is to determine the validity of LBM as a suitable alternative to other CFD methods at high Reynolds numbers the numerical domain was chosen to simulate a prior experimental set-up of which validation data was readily available \citep{Higham2001Modification,Higham-et-al-2016}. 
\\[2ex]
The experiment consists of a single obstacle placed in a water flume and the wake characteristics were measured using an acoustic doppler velocimeter (ADV) (see Fig.~\ref{FlumeFig}). The experimental condition was set such that given the obstacle with diameter, $D$, the Reynolds Number $Re_D$ was 28350.
\begin{equation}
Re_D = \frac{U_{\infty}D}{\nu}
\end{equation} 
where $U_{\infty}$ is the inlet velocity.
\\[2ex]
In order to fully understand whether LBM is a viable numerical method and can capture the multi-scale dynamics that occurs within porous obstacles, it is crucial to first simplify the problem to its basic components. The final objective, is to conduct a simulation of a flow moving over a fractal porous obstacle.
In this paper, we focus on strongly anisotropic flows past a single or porous obstacle.
For this scenario three obstacles were considered; a basic square cylinder and two porous obstacles, one with a regular arrangement and a second using a fractal geometry, see Fig.~\ref{ObstaclesFig} for the obstacle geometries. 

Both porous obstacles were designed so that their volume fraction, i.e. porosity was the same. Whilst we do not have the experimental data at this Reynolds number for the case of the square cylinder to perform a quantitative analysis there exists sufficient published data in literature to conduct a qualitative analysis. For all cases presented here the LBM software used was Palabos, developed at the University of Geneva by J. Latt and B. Chopard \citep{latt2009palabos}. For all three obstacle types, simulations were run over a range of turbulent Reynolds numbers, in all cases a three dimensional domain was constructed using D3Q19 lattices. Table~\ref{LBMSettingsTab} details the simulation parameters common to all three obstacles. 
\begin{table}
	\centering	
	\tbl{LBM setup parameters common to all three obstacles.}
	{\label{LBMSettingsTab}
	\begin{tabular}{lcccc}
		\hline
		Case												 	  & I     & II    & III   & IV  	\\
		\hline \hline
		Reynolds Number, $Re_D$  							  	  & 12352 & 24705 & 37057 & 49410  	\\
		Obstacle Diameter, $D$ $(m)$ 		 				  	  & 0.135 & 0.135 & 0.135 & 0.135  	\\
		Channel Length, $X$ $(m)$ 			 				  	  & 3.135 & 3.135 & 3.135 & 3.135  	\\
		Channel Width, $Y$ $(m)$			 			  	      & 0.486 & 0.486 & 0.486 & 0.486  	\\
		Flow Height, $Z$ $(m)$ 					 			  	  & 0.326 & 0.326 & 0.326 & 0.326  	\\
		Physical Inlet Velocity, $U_\infty$ $(ms^{-1})$ \num{e-3} & 91.5  & 183   & 275   & 366  	\\
		LBM Inlet Velocity, $U_{LBM}$ 				  		  	  & 0.050 & 0.100 & 0.100 & 0.100  	\\
		Physical Viscosity, $\nu$ $(m^2s^{-1})$  \num{e-6}		  & 1.00  & 1.00  & 1.00  & 1.00 	\\
		Acquisition Frequency, $\zeta_s$ $(Hz)$ 	 			  & 100   & 100   & 100   & 100 	\\
		Data sample length Time, $T$ $(s)$		 					  & 50.0  & 50.0  & 50.0  & 50.0  	\\
		Acquisition Start Time, $T_0$ $(s)$	 				  	  & 343   & 172   & 115   & 86  	\\
		\hline
	\end{tabular}}
\end{table}

Additionally, for each Reynolds number a mesh sensitivity analysis was conducted using the same mesh densities for all four cases. Due to differing inlet velocities for each case, variation in the timestep and subsequently the relaxation time for each mesh is to be expected, as an example Table~\ref{LBMMeshSensitivityTab} shows the corresponding parameters for case IV, $Re_D=49410$. For all other Reynolds numbers the same mesh densities were used, however, the remaining parameters will differ due to differing LBM inlet velocity.
This being a water channel, the domain boundary conditions remained the same irrespective of the obstacle and flowrate simulated. Hence, the bottom and side walls were set to a no-slip condition whilst the top boundary was set to free-slip. A uniform inlet was set on the left side of the domain and an outlet on the right. (See Fig.~\ref{FlumeFig} for a schematic of the numerical domain.) Owing to the explicit nature of the LBM it is necessary to allow the flow to develop to the stage where it is fully developed. Therefore, for all three cases the time at which data recording starts corresponds to when the flow has cycled ten times over the entire domain. Additionally, for all flowrate cases, Smagorinsky subgrid modelling \citep{smagorinsky1963general} was selected using a Smagorinsky constant $C_s=0.2$ and multiple relaxation time (MRT) dynamics implemented \citep{Humieres2002MRT,d1992generalized}. Furthermore, in order to prevent the formation of a large gradient at the inlet when starting the simulation, the inlet velocity is gradually increased over a time period equivalent to 20~000 timesteps.

\begin{table}
	\centering	
	\tbl{LBM Mesh sensitivity parameters, for $Re_D=49410$. }
	{\label{LBMMeshSensitivityTab}
	\begin{tabular}{lccc}
		\hline
		\multirow{2}{*}{Obstacle Type}				  & Porous  	 & Solid  	   & Porous  	  \\
													  & Regular (PR) & Square (SS) & Fractal (FR) \\
		\hline \hline
		Mesh Density, $NPM$ $(Nodes/m)$   	 		  & 417     	 & 417    	   & 374		  \\
		Node Spacing, $\delta_x$ $(m)$ \num{e-3}	  & 2.397   	 & 2.397  	   & 2.673        \\
		Timestep interval, $\delta_t$ $(s)$ \num{e-4} & 6.550   	 & 6.550   	   & 7.303        \\
		LBM Viscosity, $\nu_{LBM}$ \num{e-5}		  & 11.40   	 & 11.40       & 10.22        \\
		BGK Relaxation, $\tau_{BGK}$ \num{e-3} 		  & 500.34  	 & 500.34      & 500.31       \\
		\hline
	\end{tabular}}	
\end{table}

\section{Results}
\label{secresults}

\subsection{Centreline mean profiles: streamwise velocity and TKE}

We define the normalised mean streamwise velocity as follows
\begin{equation}
u^* = \frac{\overline{u}}{U_{\infty}}
\end{equation}
where $u$ is the streamwise velocity. The normalised turbulent kinetic energy (TKE) is defined as follows
\begin{equation}
k^* = \frac{1}{2} \frac{\overline{{u'}^2} + \overline{{v'}^2} + \overline{{w'}^2}}{{U_{\infty}}^2}
\end{equation}
$u'$, $v'$ and $w'$ are respectively the velocity fluction in the streamwise, spanwise and vertical direction. Averages were performed over time.

\begin{figure}[h]  
	\centering
	\includegraphics[width=0.49\textwidth]{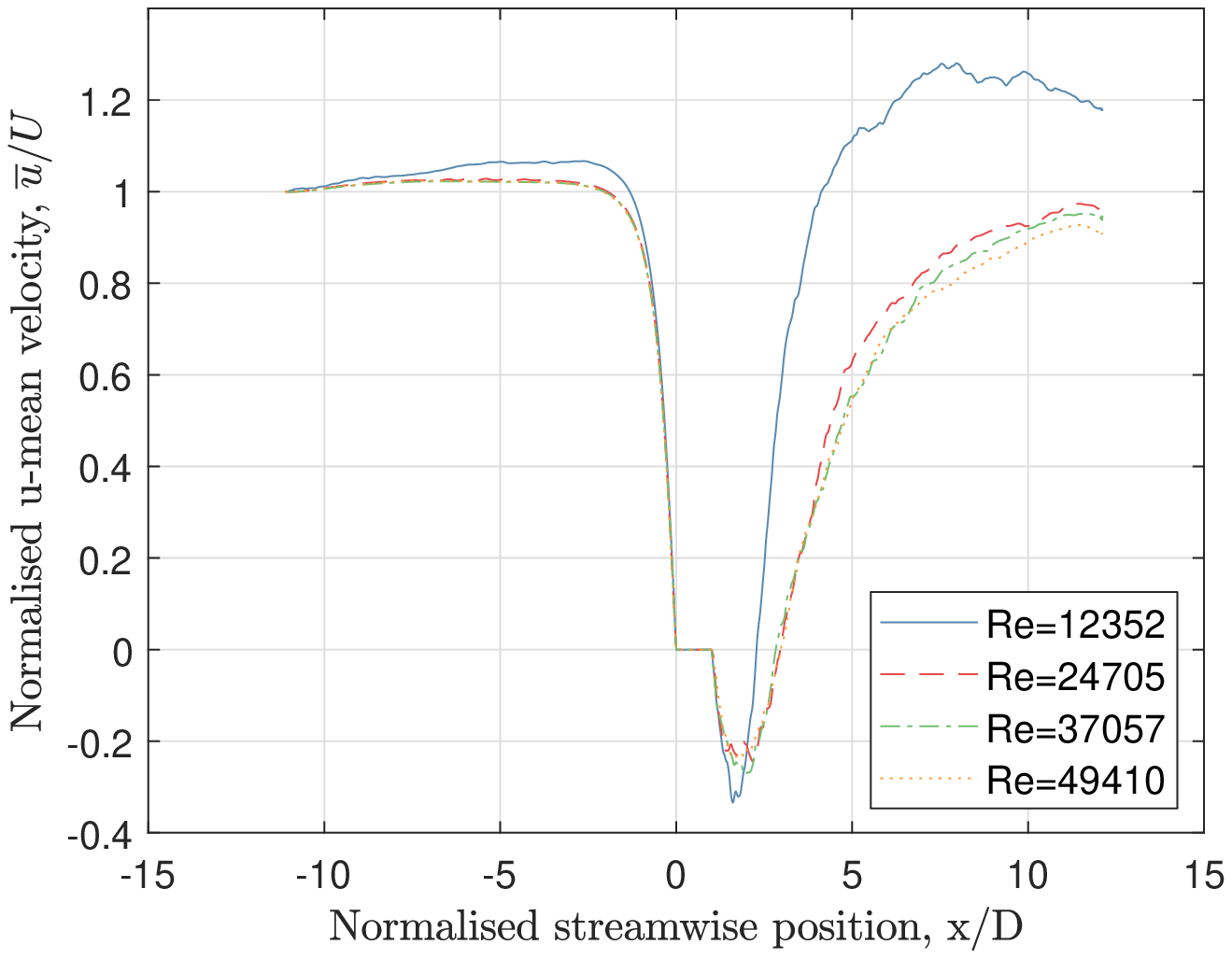}
	\includegraphics[width=0.49\textwidth]{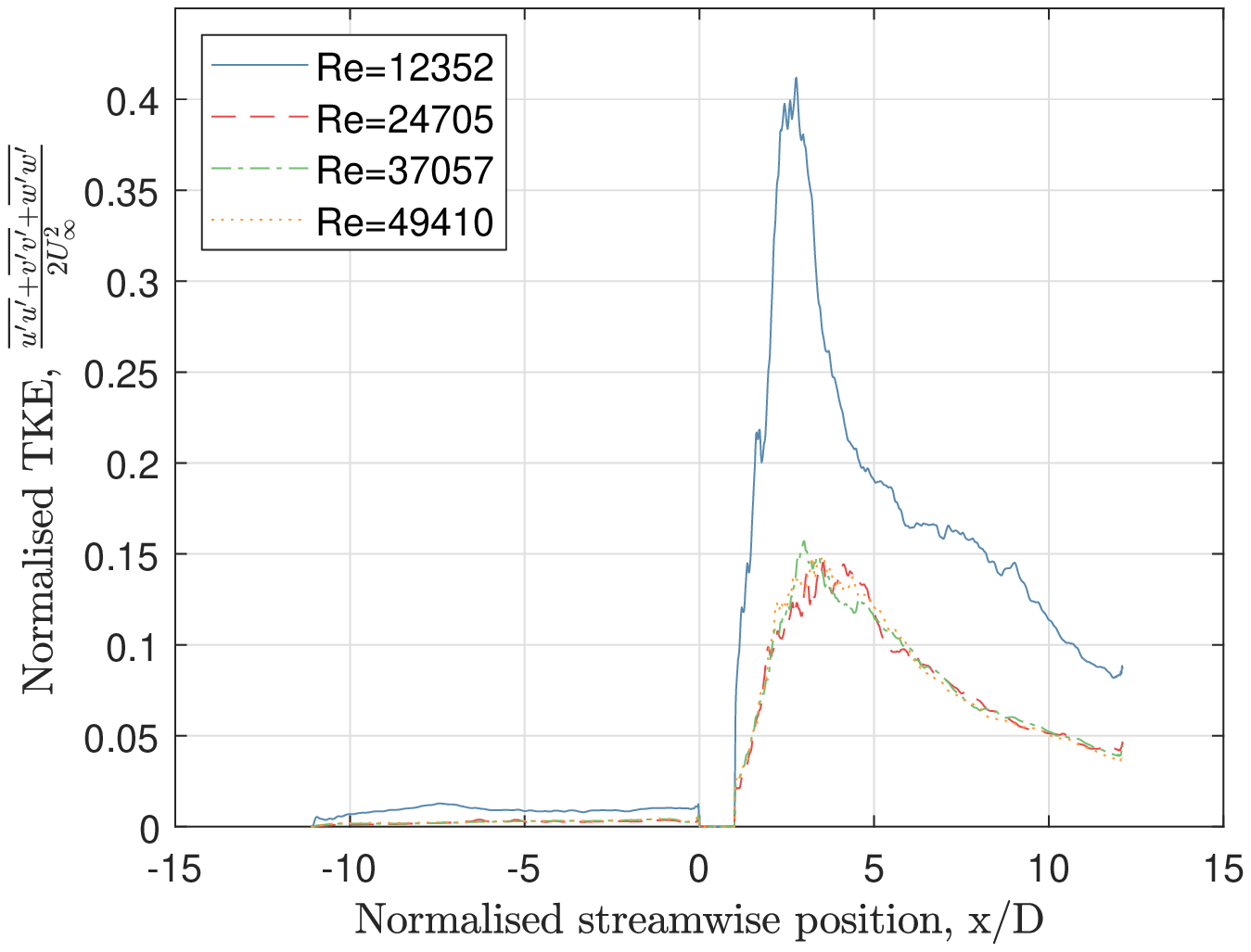}
	\caption{Mean centreline profiles for flow past a solid square cylinder in the turbulent regime. (Left) Streamwise velocity (Right) TKE}
	\label{SC_MeanU_TKE_ResultsImg}
\end{figure}

The mean streamwise velocity and turbulent kinetic energy (TKE) profiles along the centreline of the domain are given in Figs.~\ref{SC_MeanU_TKE_ResultsImg}, \ref{RG_MeanU_TKE_ResultsImg} and \ref{FC_MeanU_TKE_ResultsImg} for the solid square, regular and fractal obstacles respectively.

\subsection{Solid square cylinder (SS)}

No experimental data was available for direct comparison of the solid square obstacle case. 
So the validation of the LBM can be done on the porous cases only. It is important however to present the results for the square obstacle as well, for the sake of comparison and to understand later on the physics behind the porous obstacles' wake.

The first point to be made, is that the method appears to be more stable with increasing Reynolds numbers, which is counter intuitive as an increase in Reynolds number would mean a more complex flow, therefore, more numerically unstable. As the Reynolds number increases the normalised profiles for both mean velocity and TKE show a universal shape that is acheived after Re=24705.

The mean streamwise profiles and TKE profiles, show good agreement in both the near and far wake region in the higher flow cases with the experimental results presented in \citep{bosch1998simulation} for $Re_D=22000$ (and for $Re$ as low as 12000 in \citep{Durao-et-al-1988}) which supports the idea of reaching a universal behavious around these Reynolds numbers.

\subsection{Regular porous obstacle (PR)}

\begin{figure}[h]
	\centering
	\includegraphics[width=0.49\textwidth]{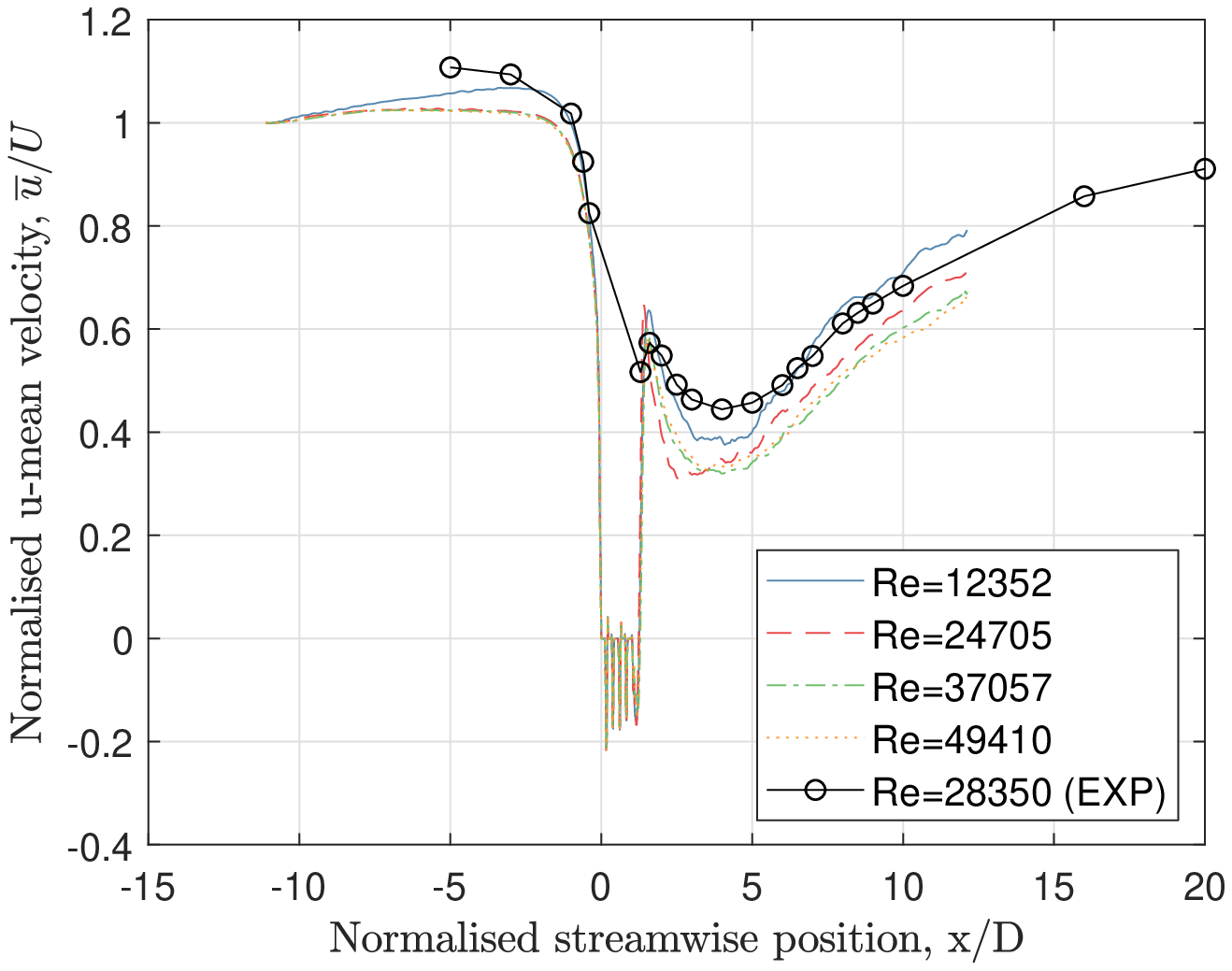}
	\includegraphics[width=0.49\textwidth]{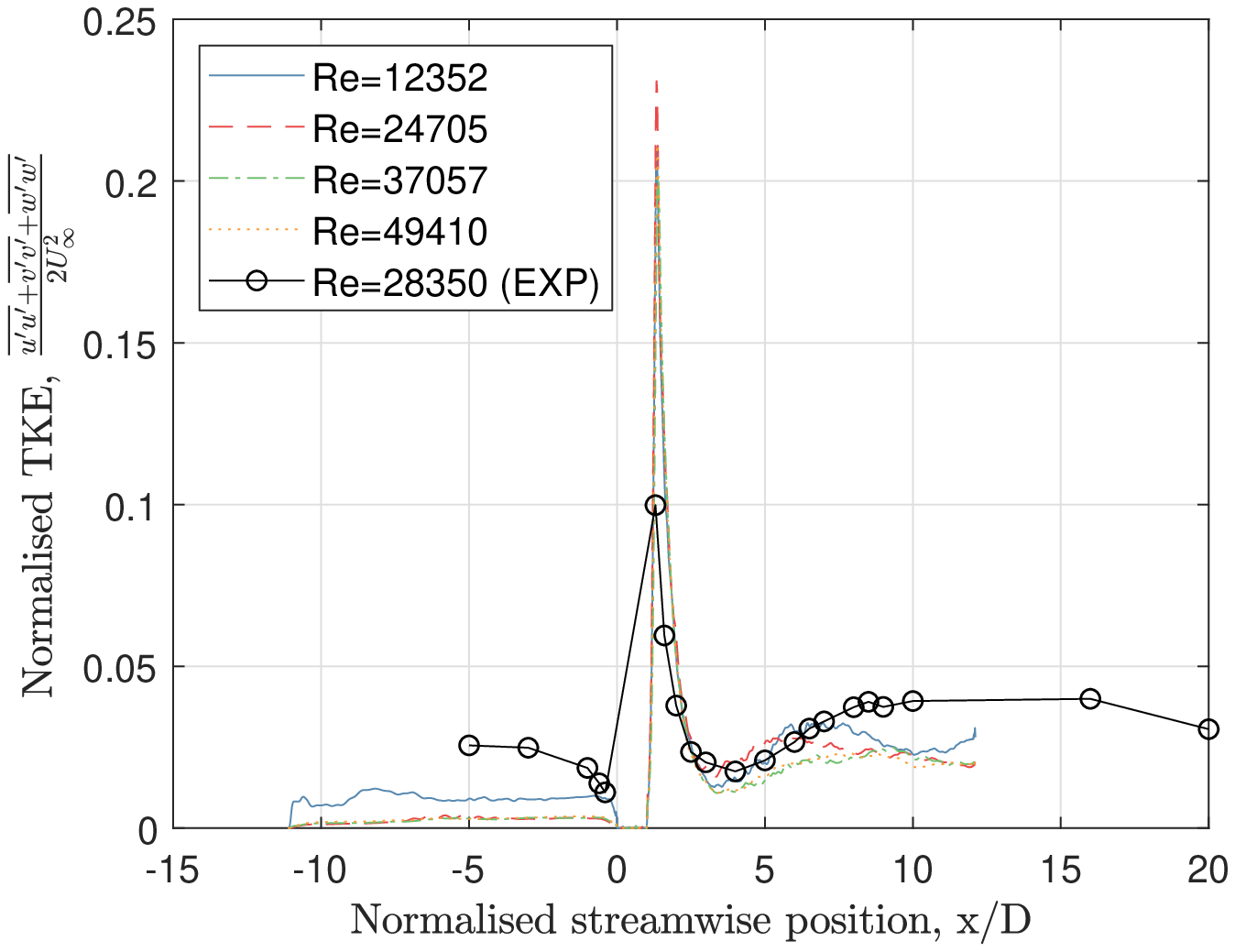}	
	\caption{Mean centreline profiles for flow past a regular obstacle in the turbulent regime. (Left) Streamwise velocity (Right) TKE}
	\label{RG_MeanU_TKE_ResultsImg}
\end{figure}
%
%
%
%
%
A similar trend to a universal behaviour is observed for the regular porous obstacle. It seems to reach it faster with less discrepency bwetween the case Re=12352 and the larger ones.

The mean velocity profiles, Fig.~\ref{RG_MeanU_TKE_ResultsImg}, show that in the case of the regular obstacle (PR) at a resolution of 417 NPM, compared to the experimental results, the LBM approximations lie below the experimental results including those of case II which is the closest match to the experimental conditions. Overall, case I is the case that best matches the experimental data, this is interesting as it is case II which most closely matches the experimental conditions, case I being less than half of the experimental Reynolds number. However, all cases remain in good agreement with the experimental results.

\subsection{Fractal porous obstacle (PF)}

\begin{figure}[h]
	\centering	
	\includegraphics[height=6.0cm]{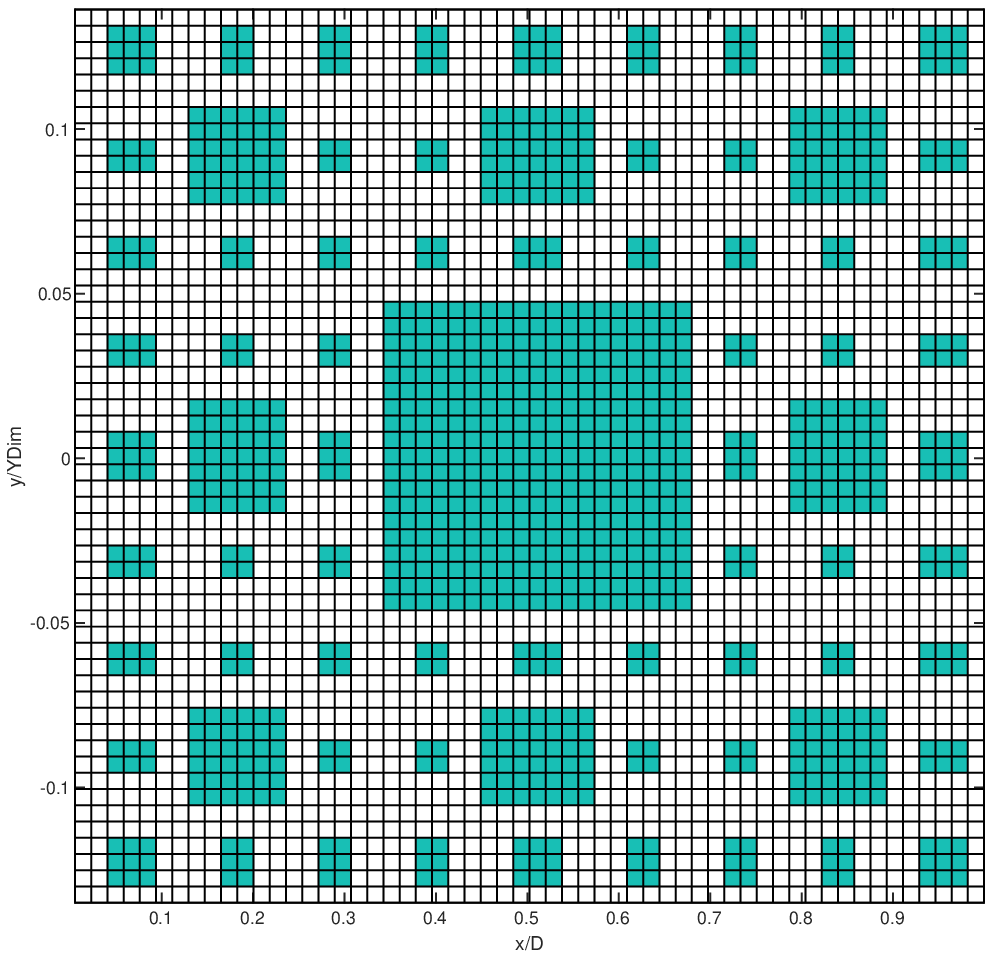}
	\includegraphics[height=6.0cm]{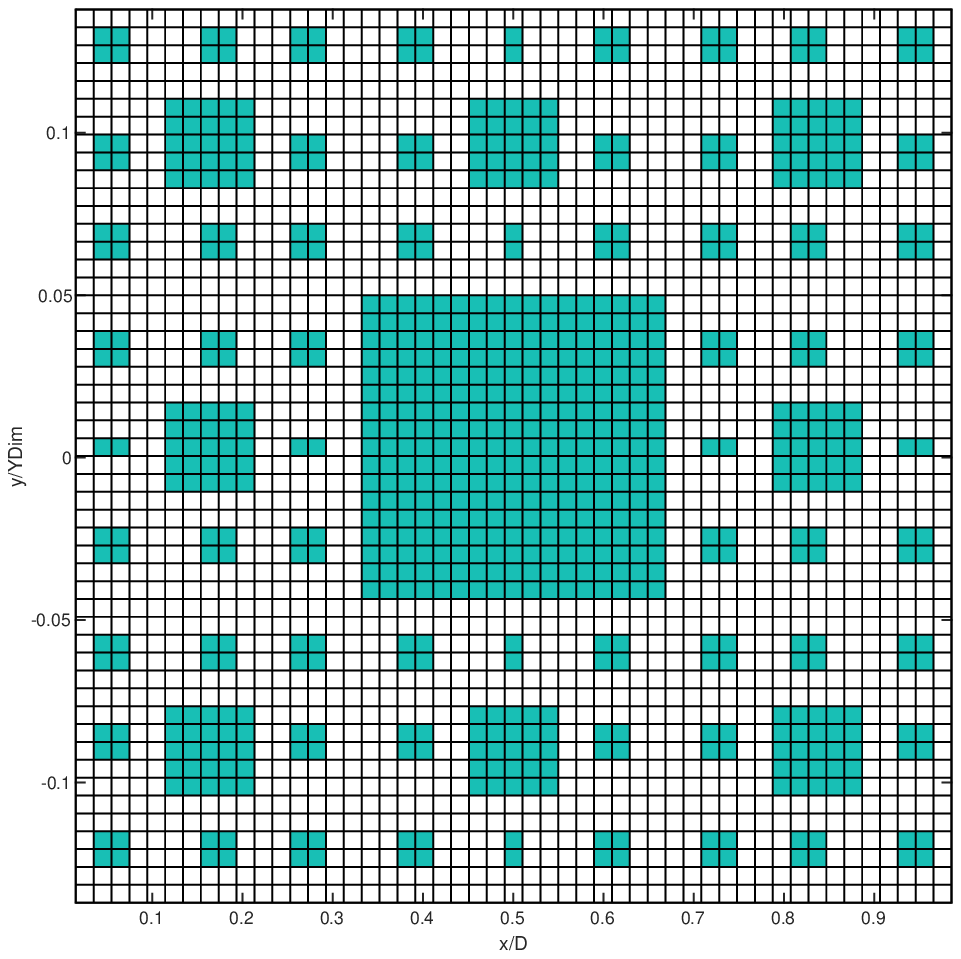}	
	\caption{Meshing of the fractal obstacle. Each square represents a node in the lattice, with the filled squares representing the obstacle. (a) 417 NPM (b) 374 NPM}
	\label{FC_MeshImg}
\end{figure}
In the case of the fractal obstacle, comparisons using the highest resolution, 417 NPM, are not appropriate. From a mesh analysis the mesh of the fractal obstacle at the highest resolution, Fig.~\ref{FC_MeshImg}, shows that although the lattice spacing is smaller than the smallest iteration of the fractal geometry, not all instances of the third iteration obstacles have the same size for the 417 NPM resolution, however, a 374 NPM resolution is a more faithful representation of the fractal geometry as the majority of the individual obstacles maintain the square cross-section. This fact is quite substantial as it indicates that the flow is heavily influenced by the geometry of the fractal, which would lead to the speculation that the same obstacle but with the sub-obstacles arranged in a different manner would yield an entirely different near wake.
\\[2ex]
\begin{figure}[h]
	\centering
	\includegraphics[width=0.49\textwidth]{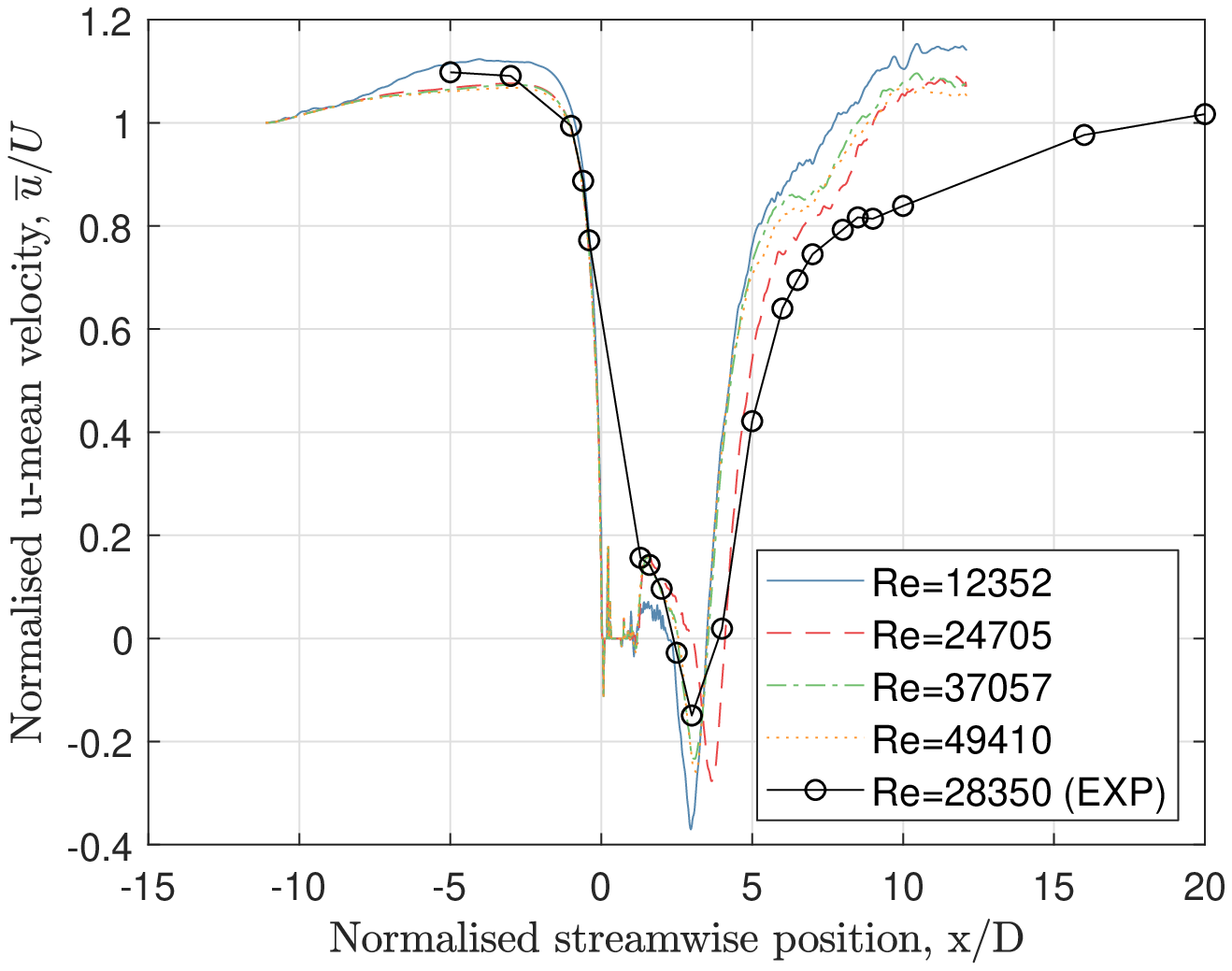}
	\includegraphics[width=0.49\textwidth]{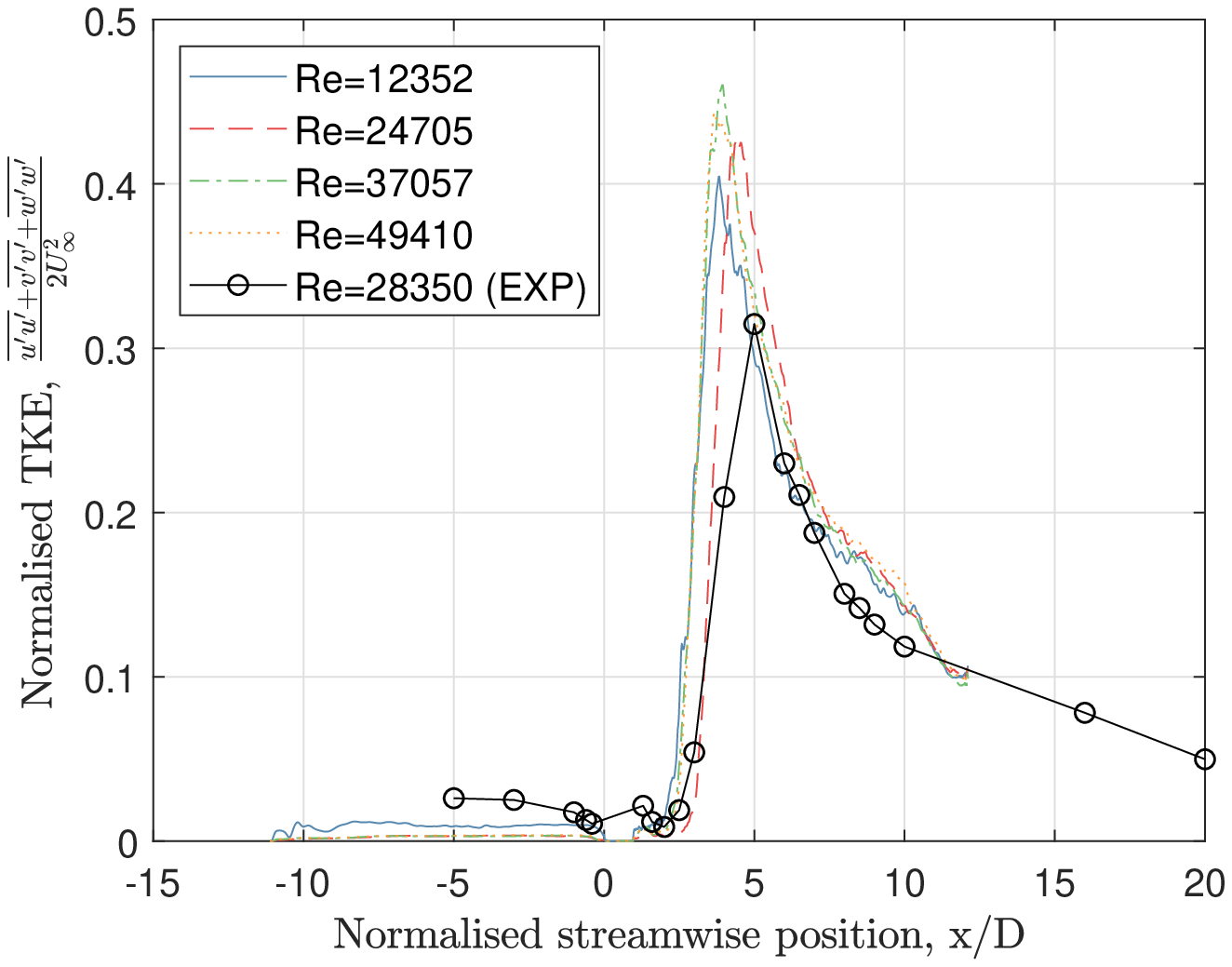}
	\caption{Mean centreline profiles for flow past a fractal obstacle in the turbulent regime. (Left) Streamwise velocity (Right) TKE}
	\label{FC_MeanU_TKE_ResultsImg}
\end{figure}
Returning to the streamwise profiles of Figs.~\ref{RG_MeanU_TKE_ResultsImg} and \ref{FC_MeanU_TKE_ResultsImg}, the acceleration immediately after the obstacle is well captured by the LBM for cases II-IV, however, case I does demonstrate this effect to a lesser extent. Given that case I represent flow speeds slower than the experimental data it could be that this behaviour is specific to the higher flowrate cases. Subsequently, there is a disagreement between the LBM cases for the location of the profile minima, with cases I, III, IV predicting a location closer to the obstacle than case II. Although this location predicted by the first group appears to agree with the experimental data it may not be correct. Given the undersampling of the velocity profile in this region, if one were to interpolate the experimental data using the minima of the profile predicted by the LBM, the location of the minima would agree more with case II. 

In the far wake region the LBM results all show significant disagreement with the experimental results, with all cases showing an unphysical flow acceleration close to the outlet. This is to be expected as it is where the subgrid model would struggle the most to match the complex vortex shredding interaction occurring there. The simple Smagorinski closure we used is only valid for homogeneous isotropic turbulence cascade and clearly breaks down in this region.

Looking at the TKE profiles, in the case of the regular obstacle there is significant agreement between the LBM results and the experimental data. Since the experimental data was collected via an ADV, which is an intrusive method data near or close to the obstacle is very difficult to collect, and since the peak occurs immediately aft of the obstacle its natural for the experimental data in this region to be underestimated. 

For the fractal obstacle, the TKE profiles, show a decent agreement with experimental data, more so in the far wake than in the near wake. In the near wake region, it is more likely that the peak TKE was not correctly captured in the experimental data due to under sampling along the profile.

\subsection{Transverse mean profiles: streamwise velocity}

\begin{figure}[ht]
	\centering
	\makebox[0.66\textwidth][s]{SS  PR  PF}	
	\par\medskip		
	\includegraphics[width=0.32\textwidth]{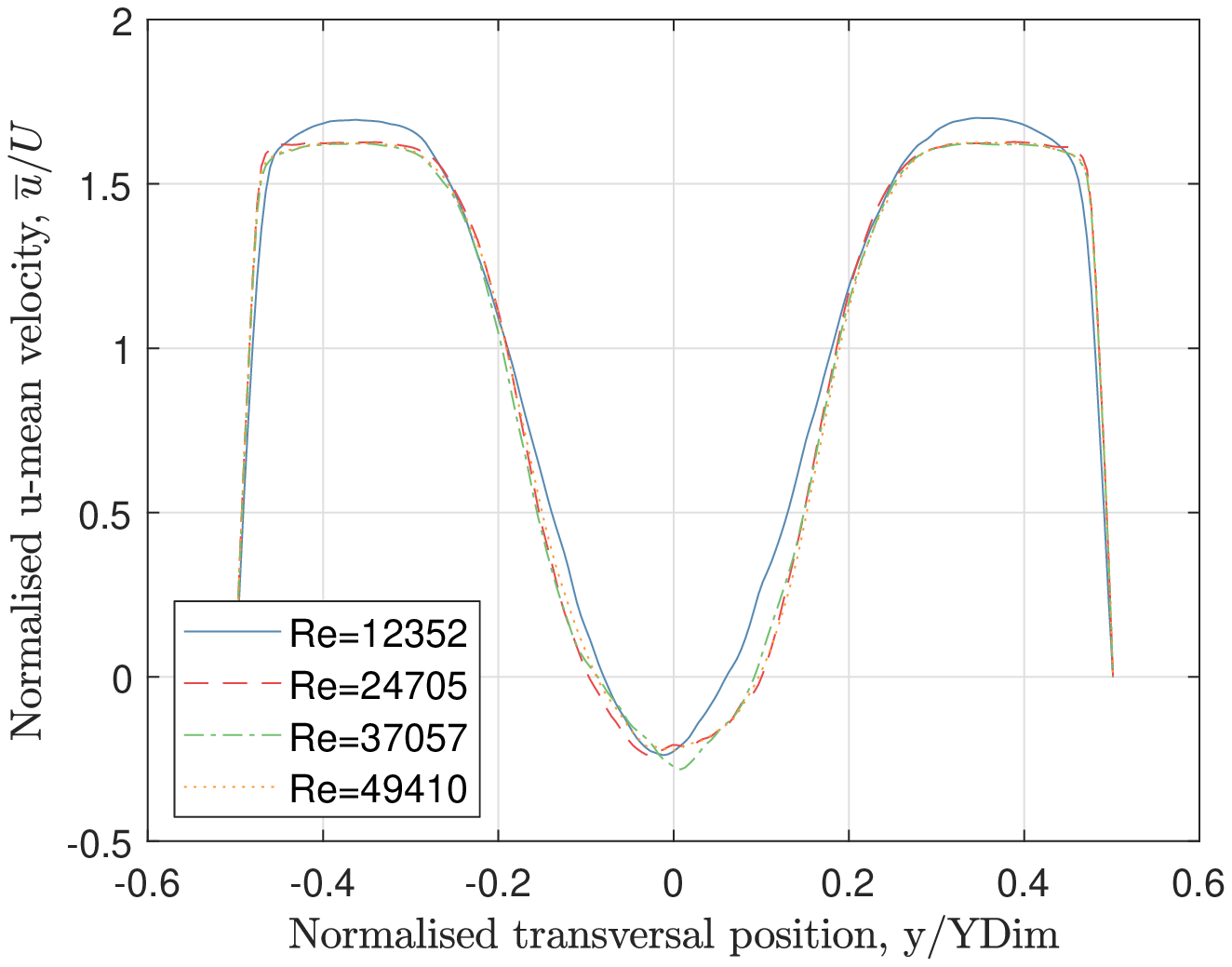}	
	\includegraphics[width=0.32\textwidth]{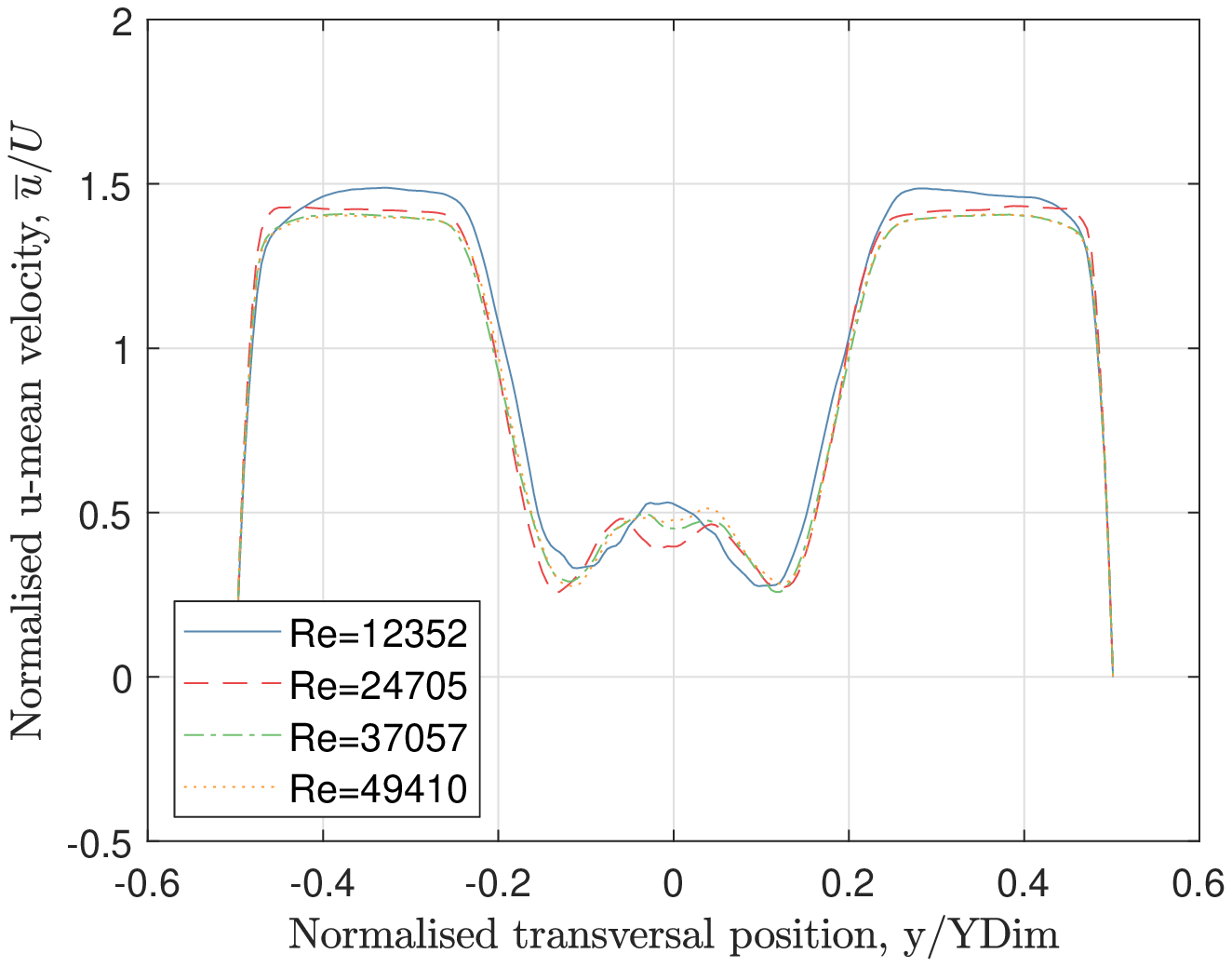}
	\includegraphics[width=0.32\textwidth]{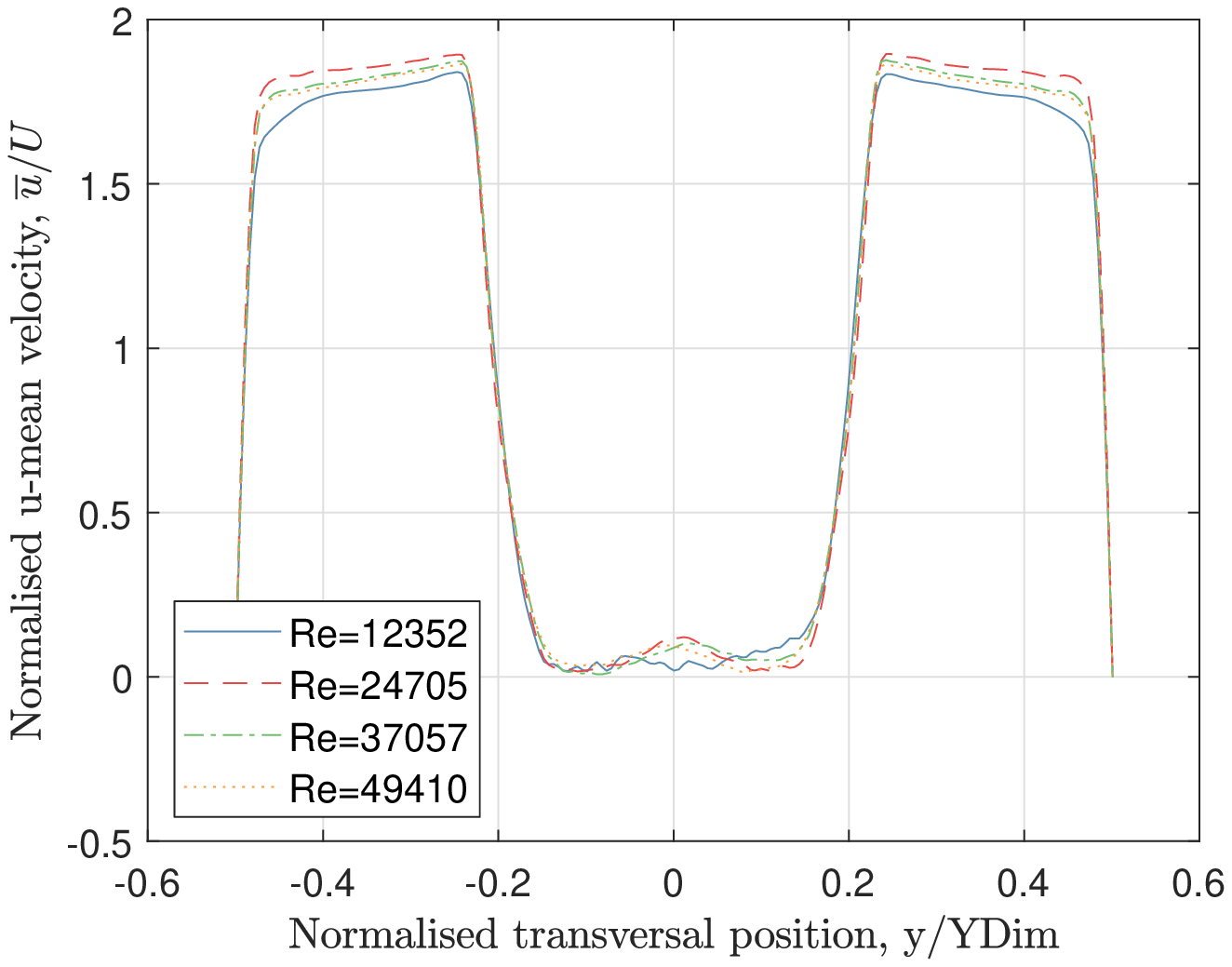}
	\par\medskip	
	\includegraphics[width=0.32\textwidth]{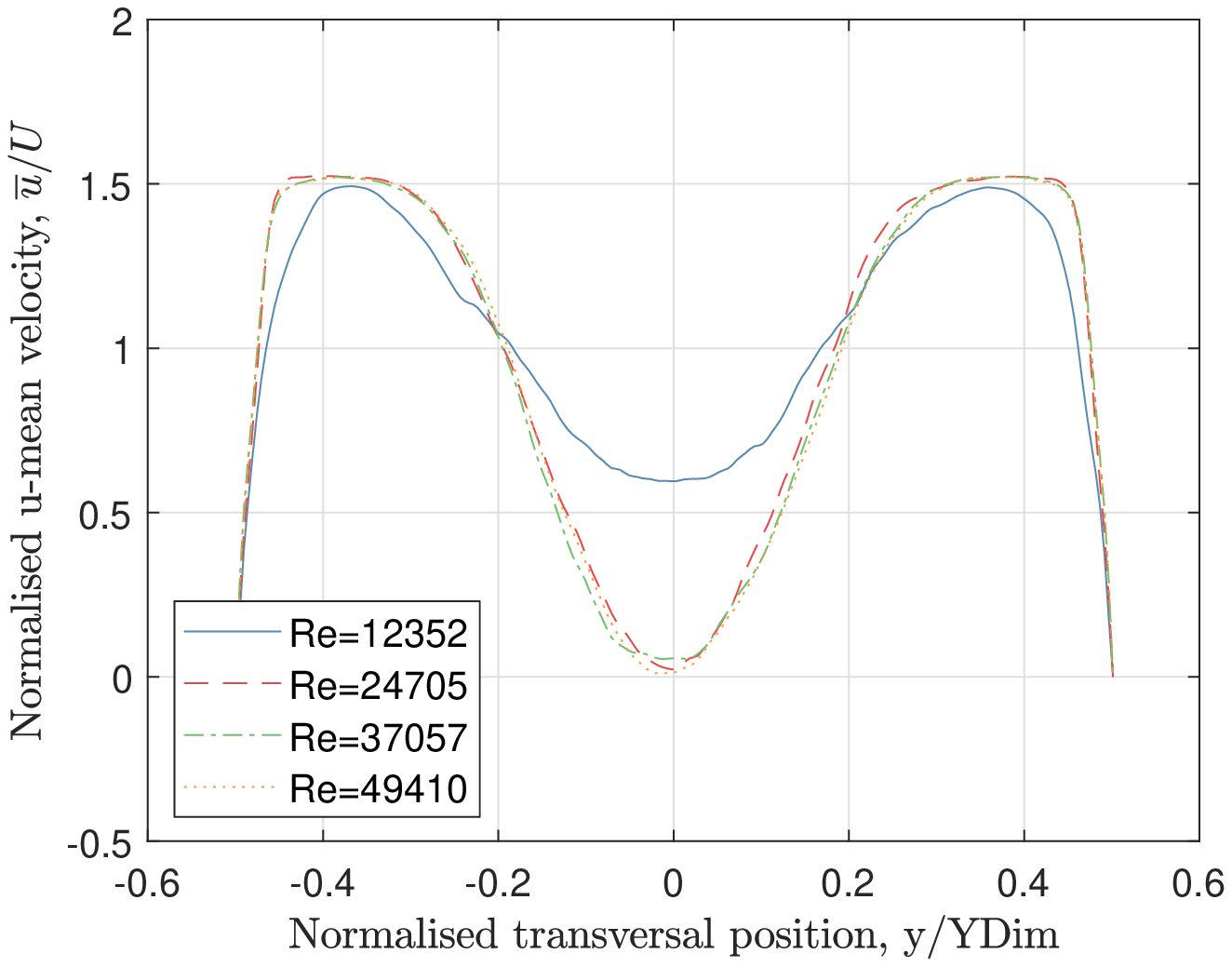}	
	\includegraphics[width=0.32\textwidth]{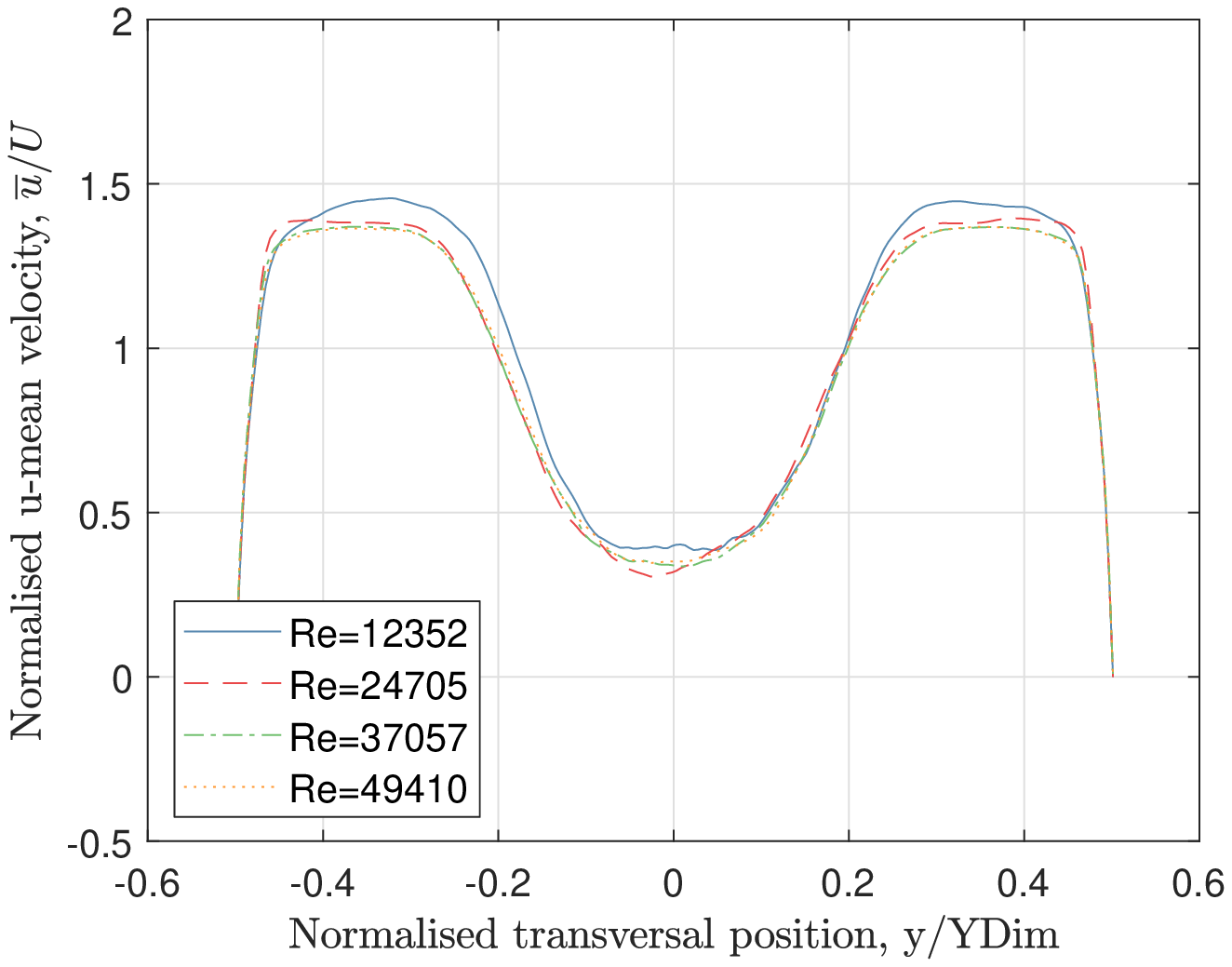}
	\includegraphics[width=0.32\textwidth]{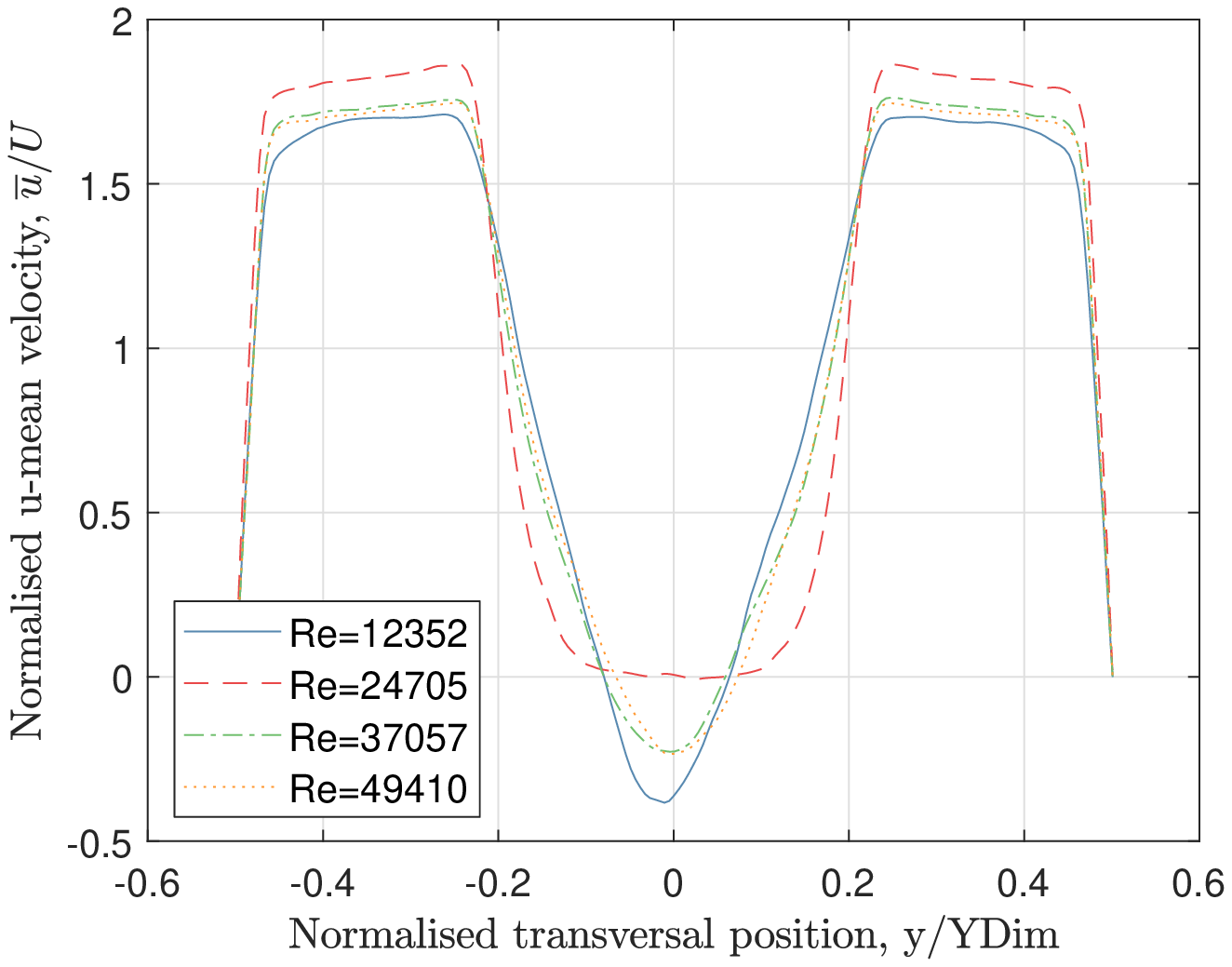}
	\par\medskip	
	\includegraphics[width=0.32\textwidth]{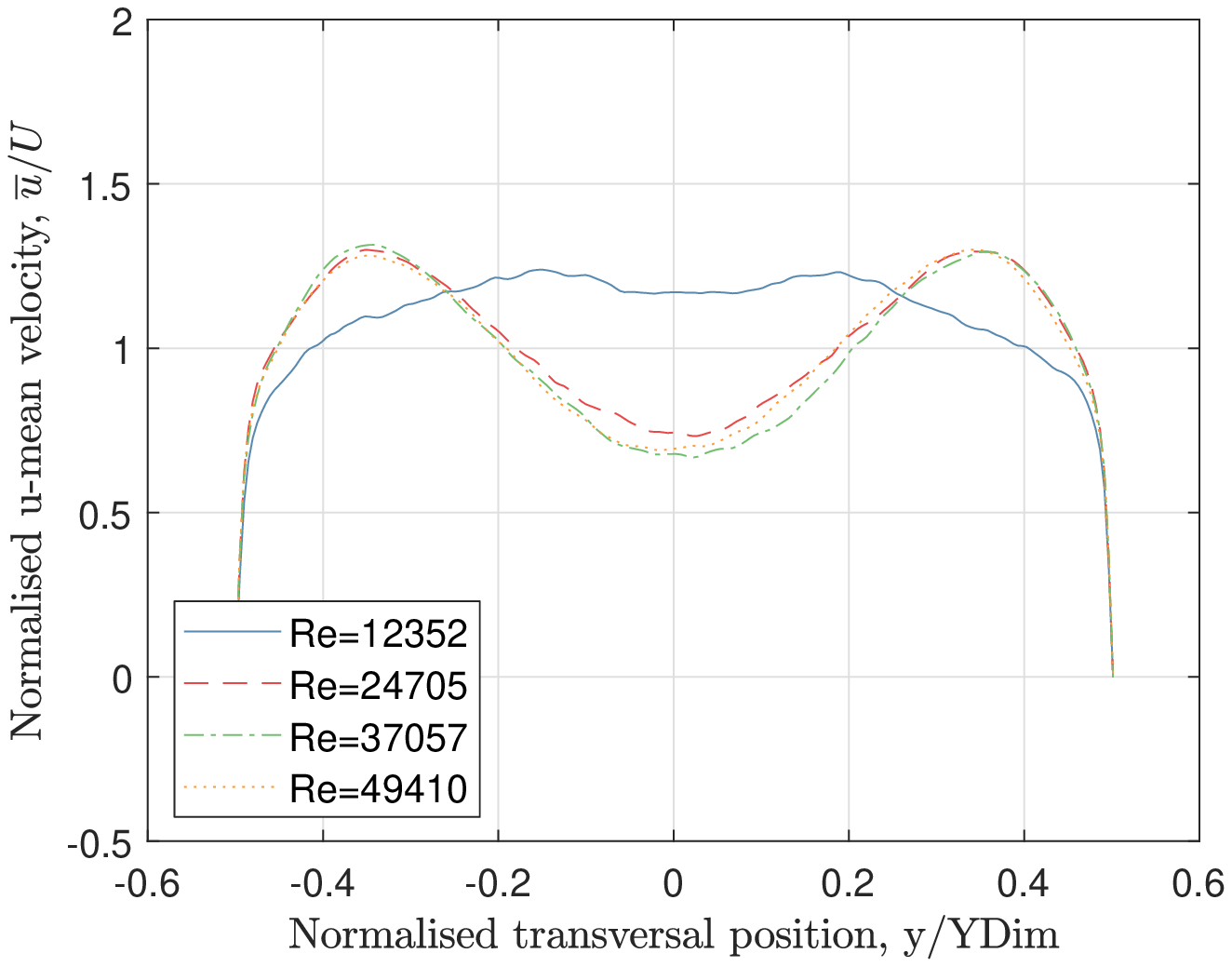}	
	\includegraphics[width=0.32\textwidth]{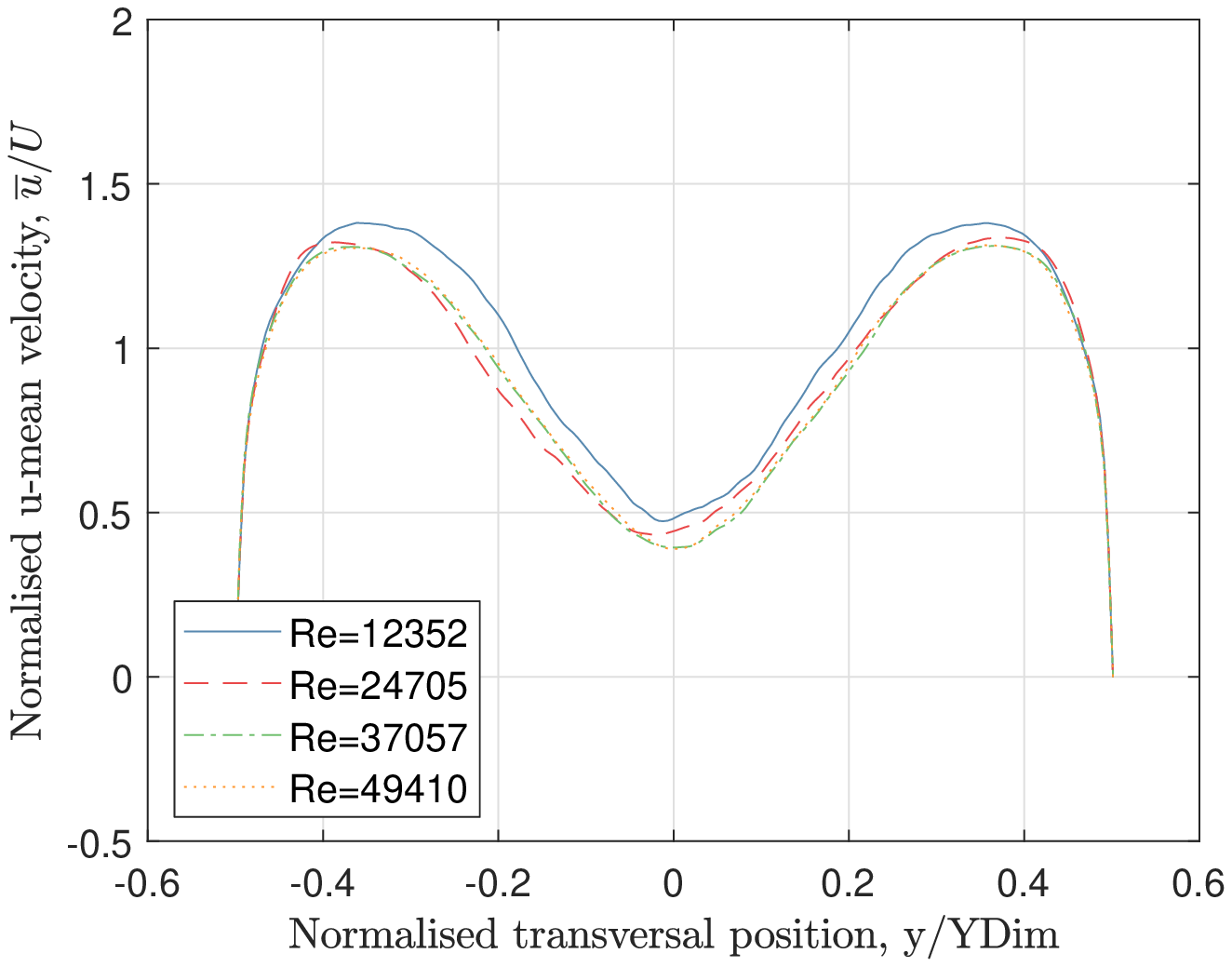}
	\includegraphics[width=0.32\textwidth]{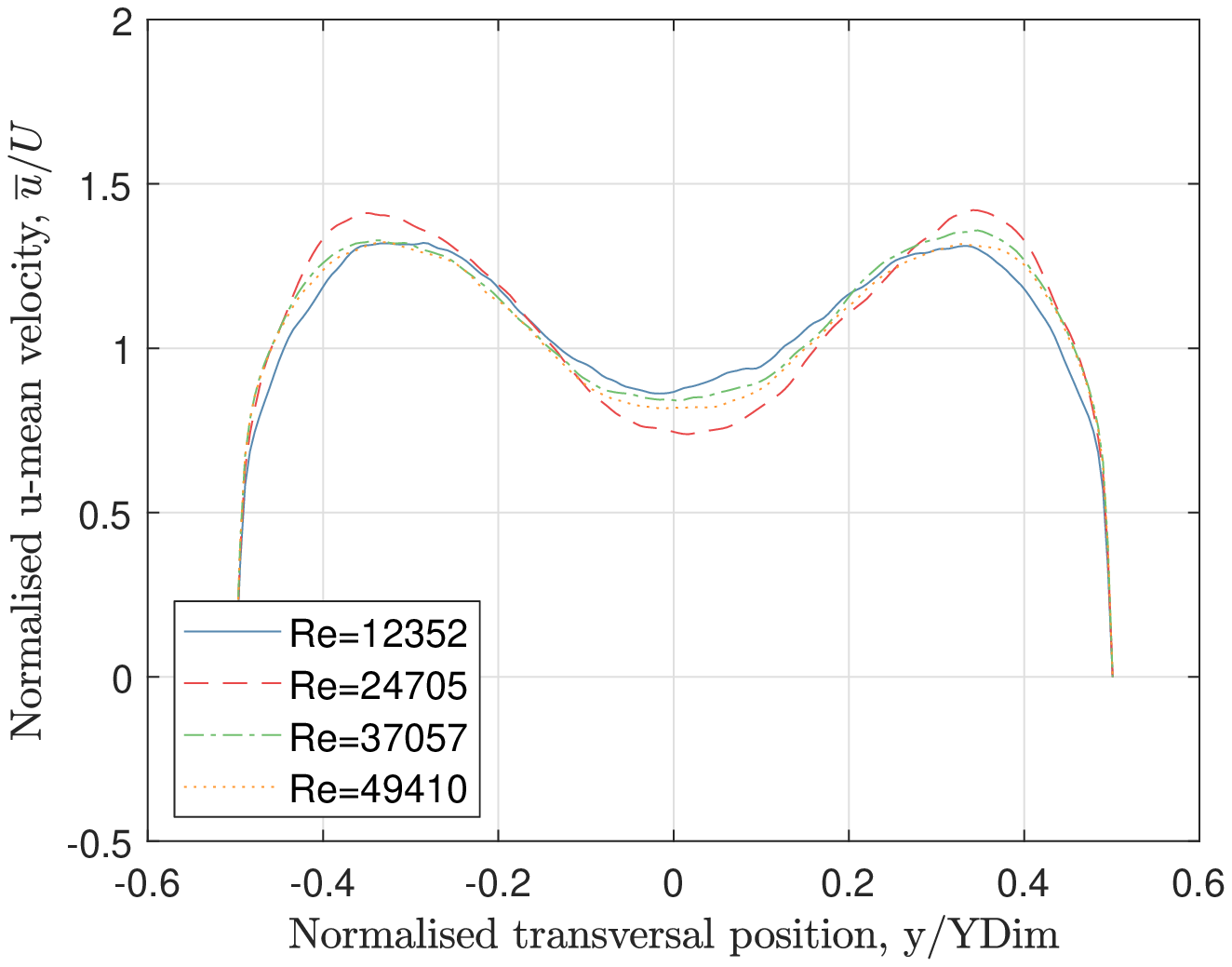}
	\par\medskip	
	\includegraphics[width=0.32\textwidth]{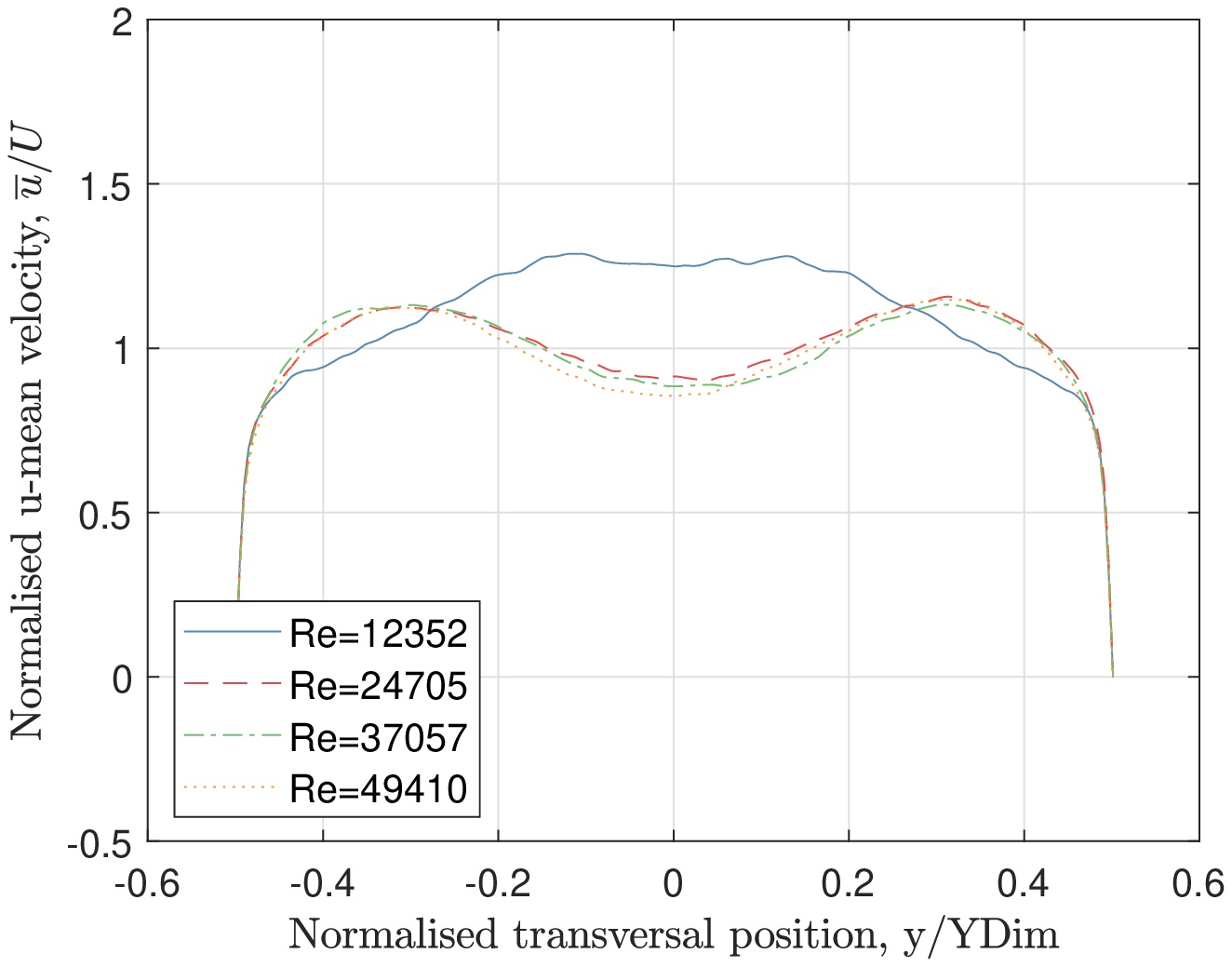}	
	\includegraphics[width=0.32\textwidth]{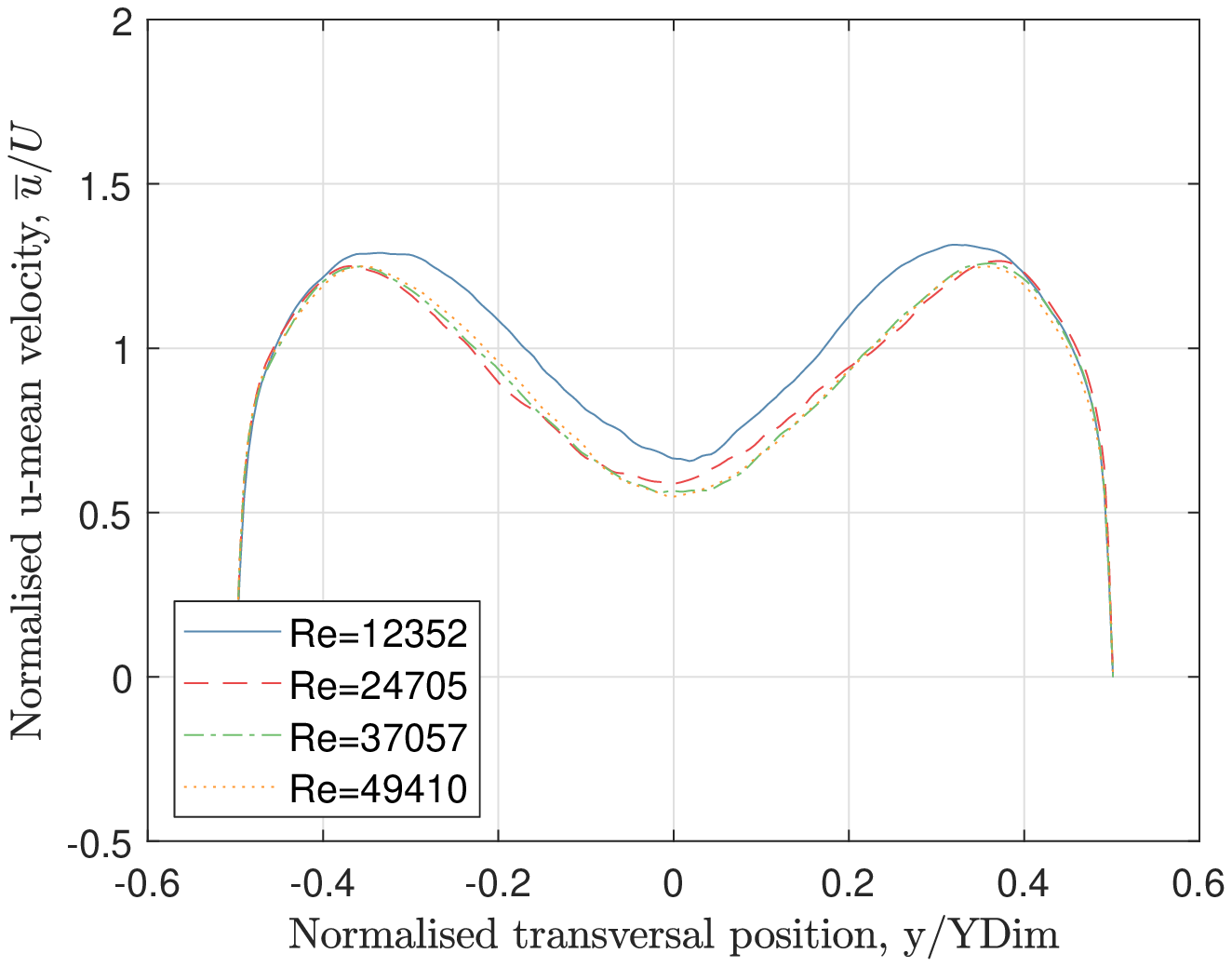}
	\includegraphics[width=0.32\textwidth]{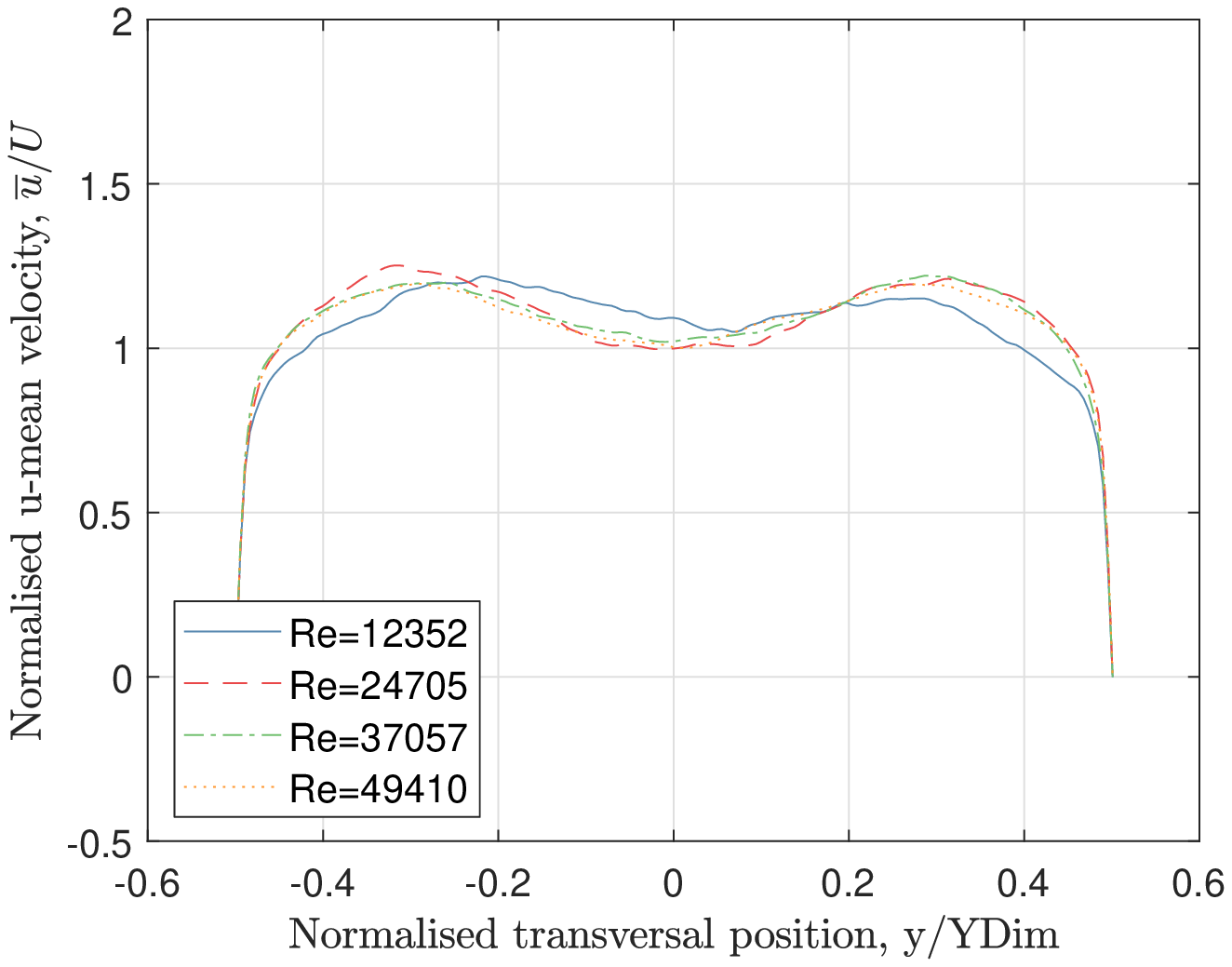}
	\par\medskip	
	\includegraphics[width=0.32\textwidth]{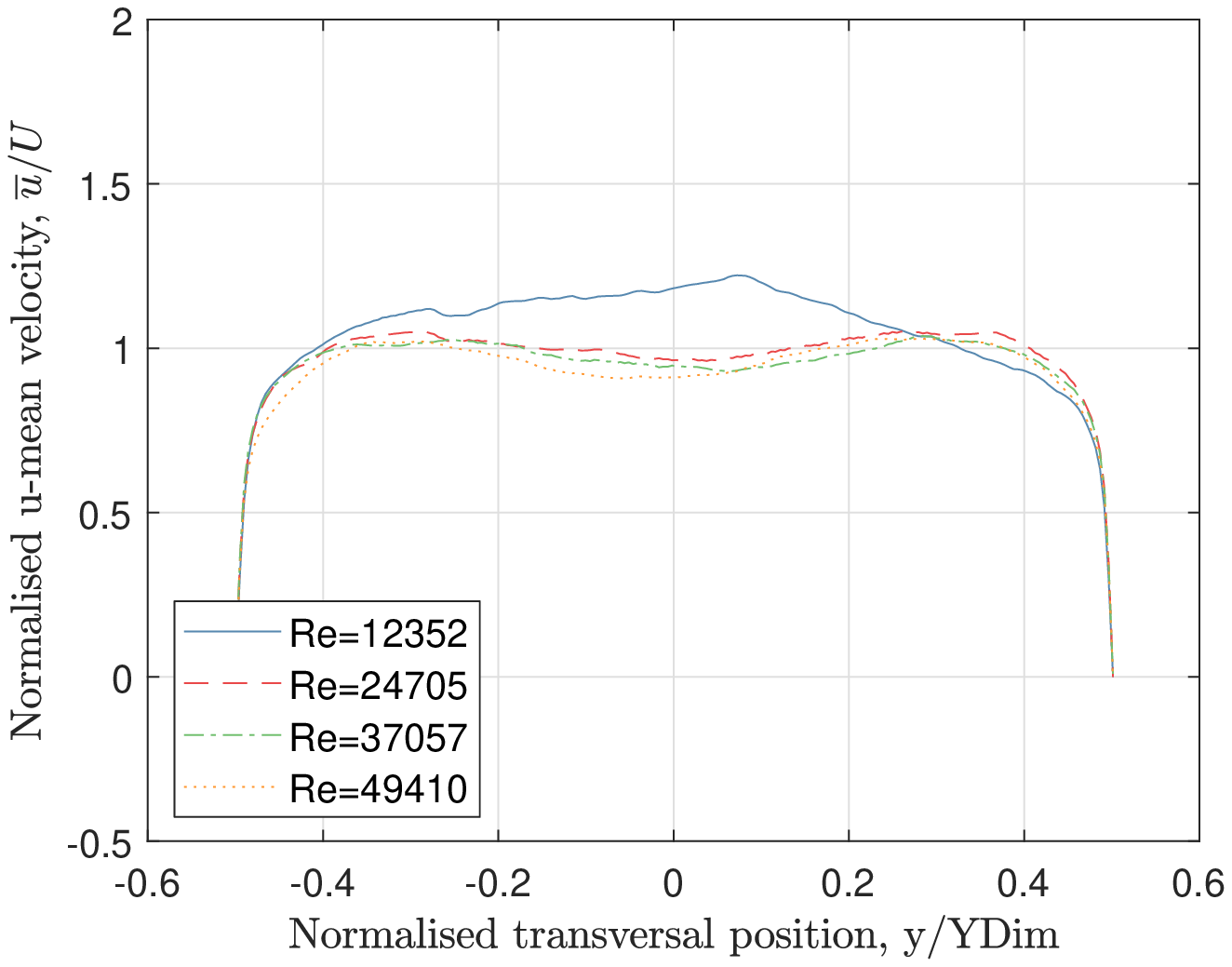}	
	\includegraphics[width=0.32\textwidth]{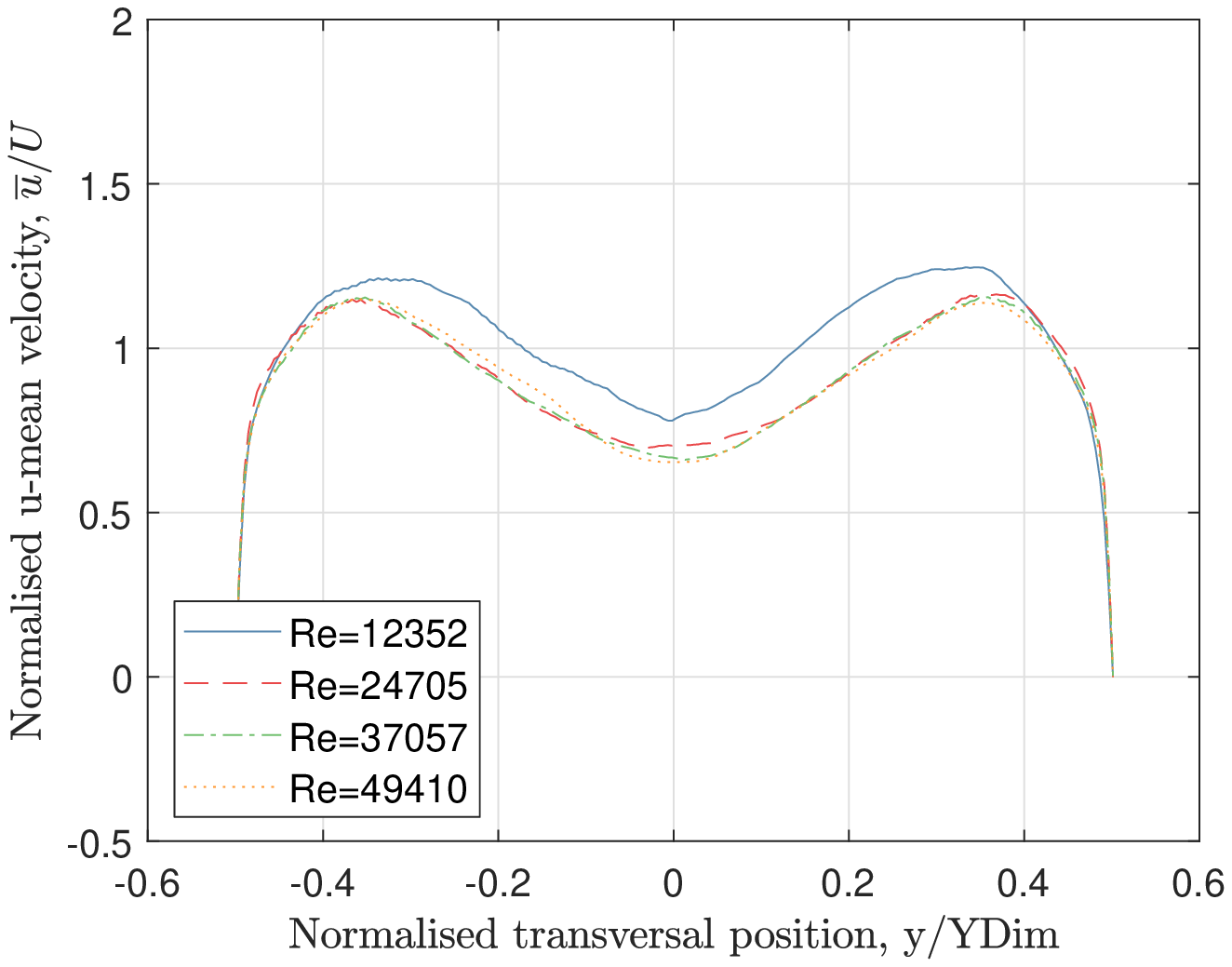}
	\includegraphics[width=0.32\textwidth]{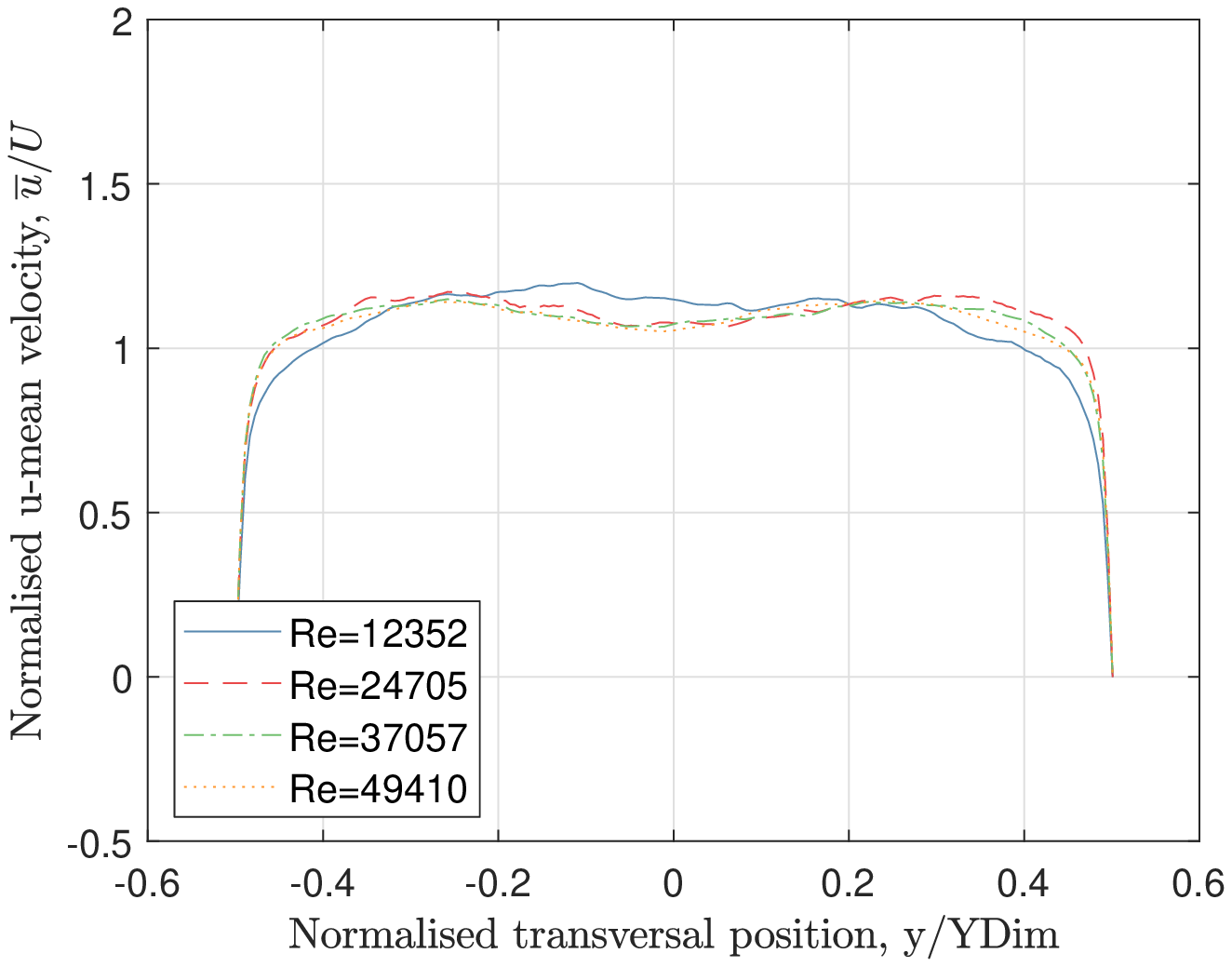}
	\caption{Transverse streamwise velocity profiles at (from top to bottom) 2D, 3D, 6D, 9D and 12D.}
	\label{MRT_Turbulent_MeanU_TransProfileImg}
\end{figure}

The evolution of the wake can further be observed in the transverse direction. 
The transverse streamwise velocity profiles for all three obstacles (SS - Solid Square, PR - Porous Regular, PF - Porous Fractal) are shown in Fig.~\ref{MRT_Turbulent_MeanU_TransProfileImg}. Profiles are taken at different locations in the channel. Each row represents a streamwise position after the obstacle, these are (from top to bottom) 2D, 3D, 6D, 9D and 12D respectively.
%
%
\\[2ex]
Both the streamwise, Fig.~\ref{MRT_Turbulent_MeanU_TransProfileImg}, and cross-stream, Fig.~\ref{MRT_Turbulent_MeanV_TransProfileImg}, velocity profiles in this direction help to characterise the shape of the wake for each obstacle. At 1D length after the obstacle we can observe three distinct wake shapes for each obstacle. Given that the regular obstacle allows the flow to pass unimpeded on alternate rows, the retardation of the velocity is significantly less than the other obstacles with the central gap acting as a small nozzle. Furthermore, the effects of the obstacle can clearly be seen longer into the far wake than the solid and fractal obstacles. For the fractal case although it can be seen that by 6D the wake is approximately the same as the solid square obstacle for cases II-IV considering it took the square obstacle 5D to reach to this point from a recirculating wake it took the fractal obstacle 3D. Furthermore, in the far wake, for the fractal the flow is less affected by the obstacle compared to solid case with increasing Reynolds numbers. Another evidence that flows through fractal objects relax faster to the fully developped turbulence pattern \citep{Nicolleau-et-al-JOT-2011}.

Furthermore, it can be noted that for the smallest velocity (case I) the flow around the solid obstacle recovers much faster than for the other cases. This contrasts with the porous obstacles, wherein all cases recover at approximately the same rate. This confirms they have reached a universal Reynolds number independent behaviour.


\subsection{Transverse mean profiles: spanwise velocity}

\begin{figure}[ht]   
	\centering
	\makebox[0.66\textwidth][s]{SS  PR  PF}	
	\par\medskip		
	\includegraphics[width=0.32\textwidth]{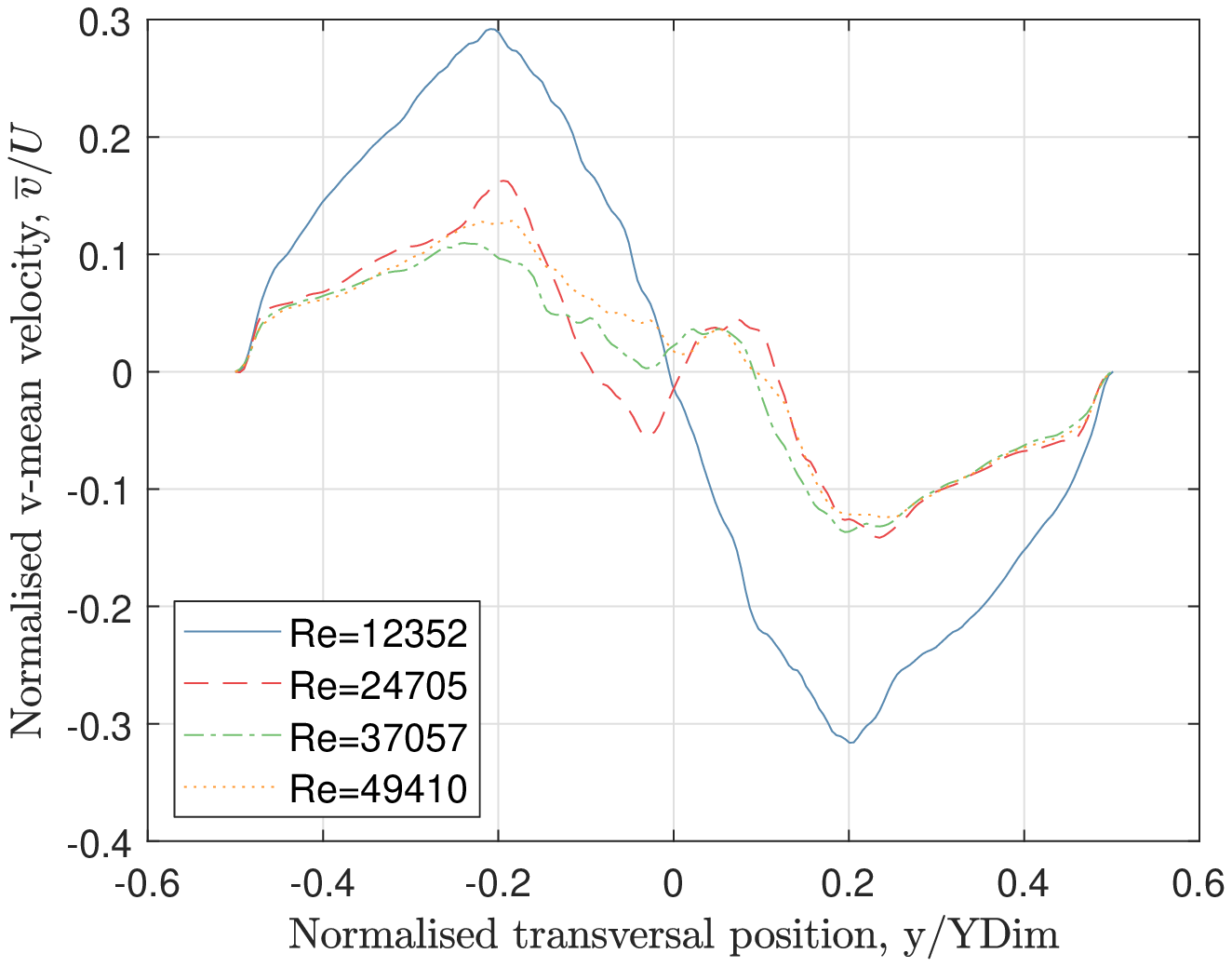}	
	\includegraphics[width=0.32\textwidth]{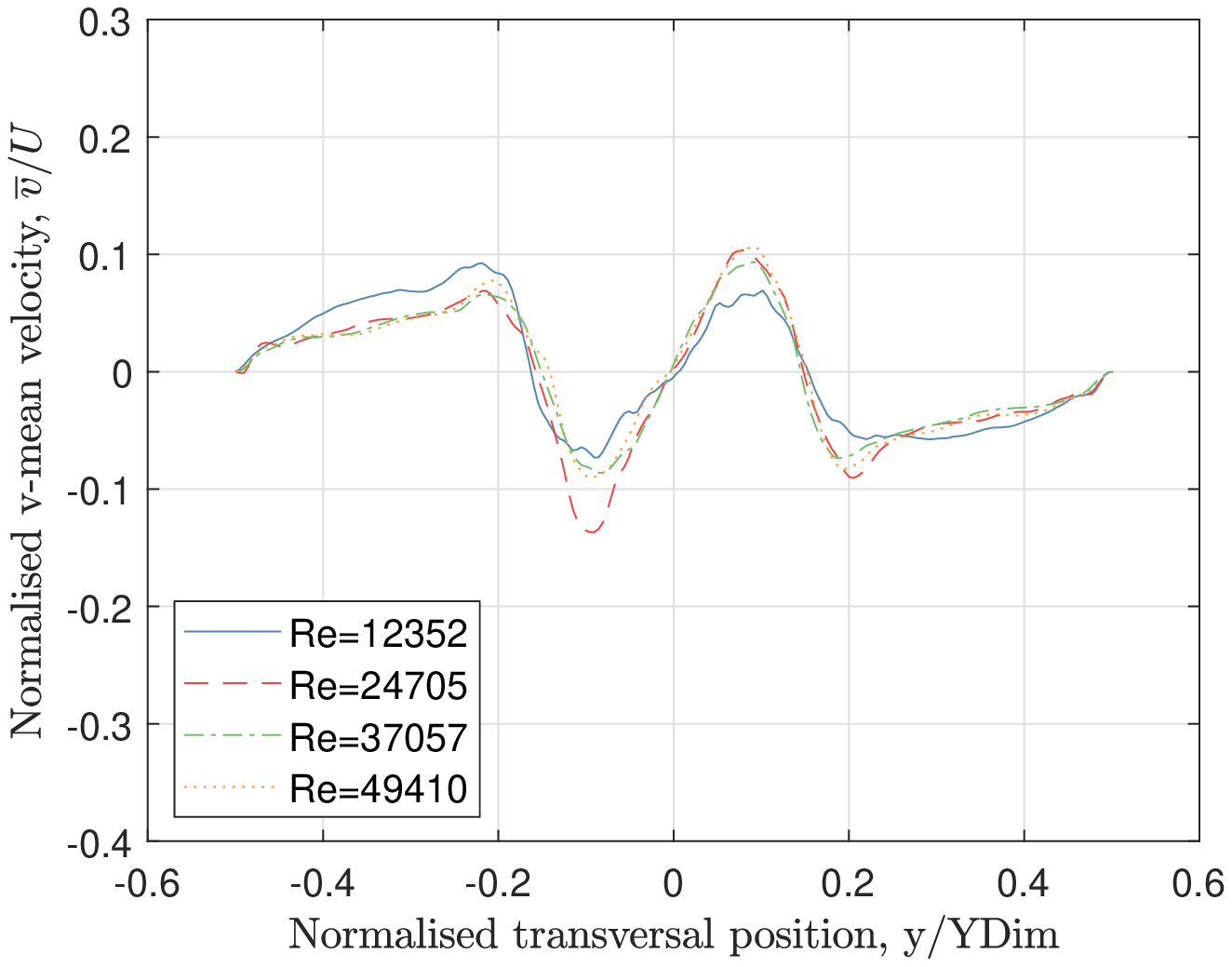}	
	\includegraphics[width=0.32\textwidth]{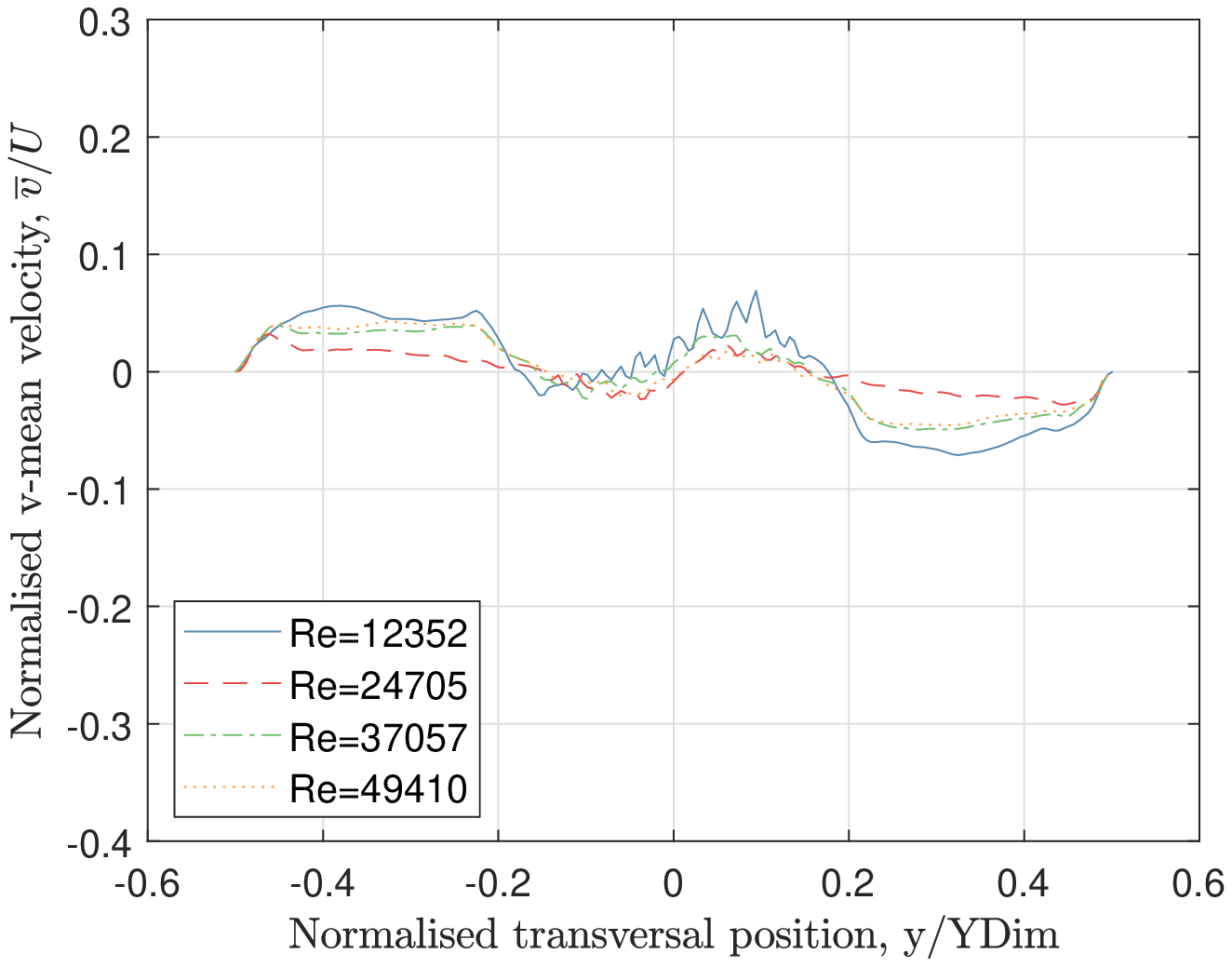}	
	\par\medskip			
	\includegraphics[width=0.32\textwidth]{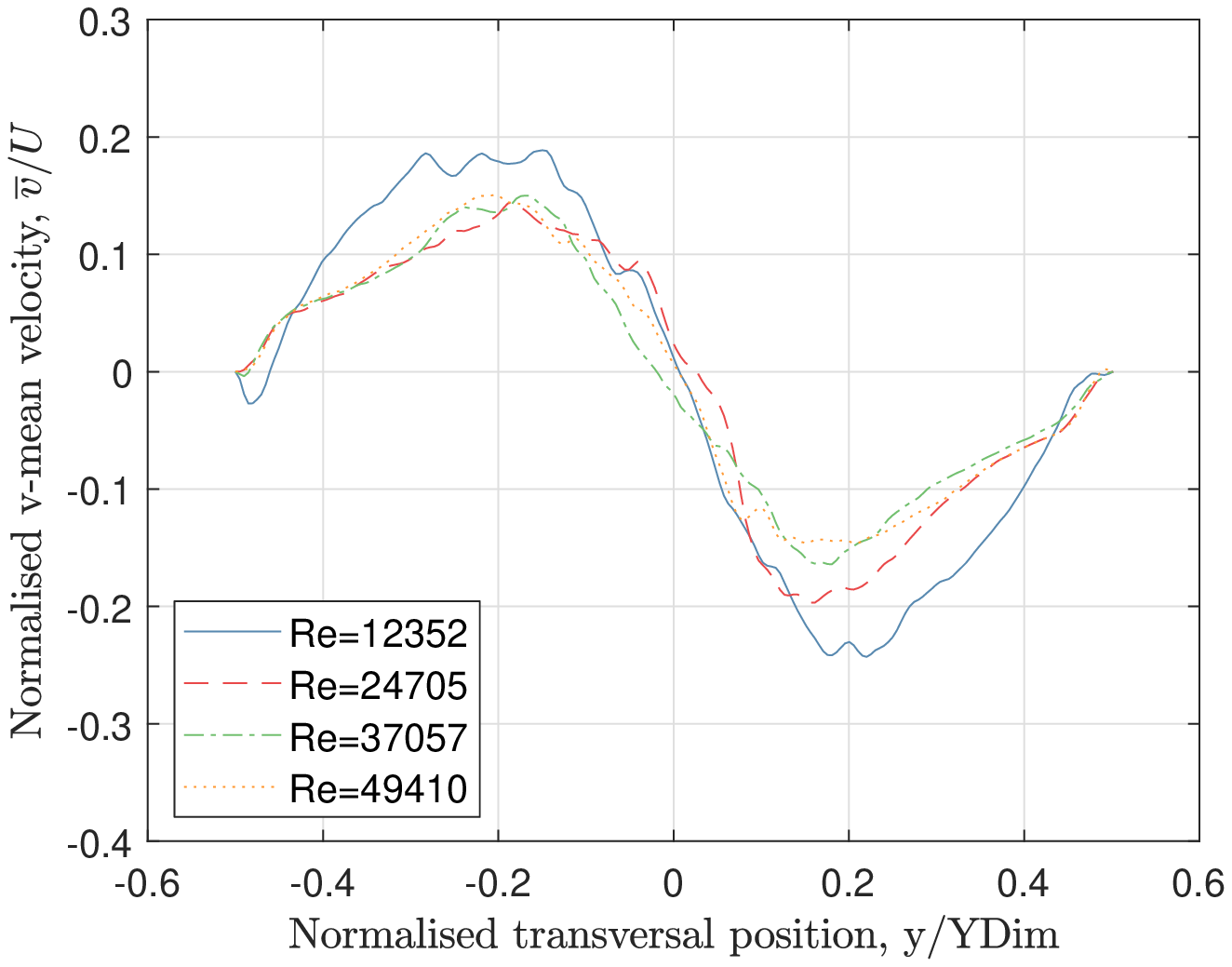}	
    \includegraphics[width=0.32\textwidth]{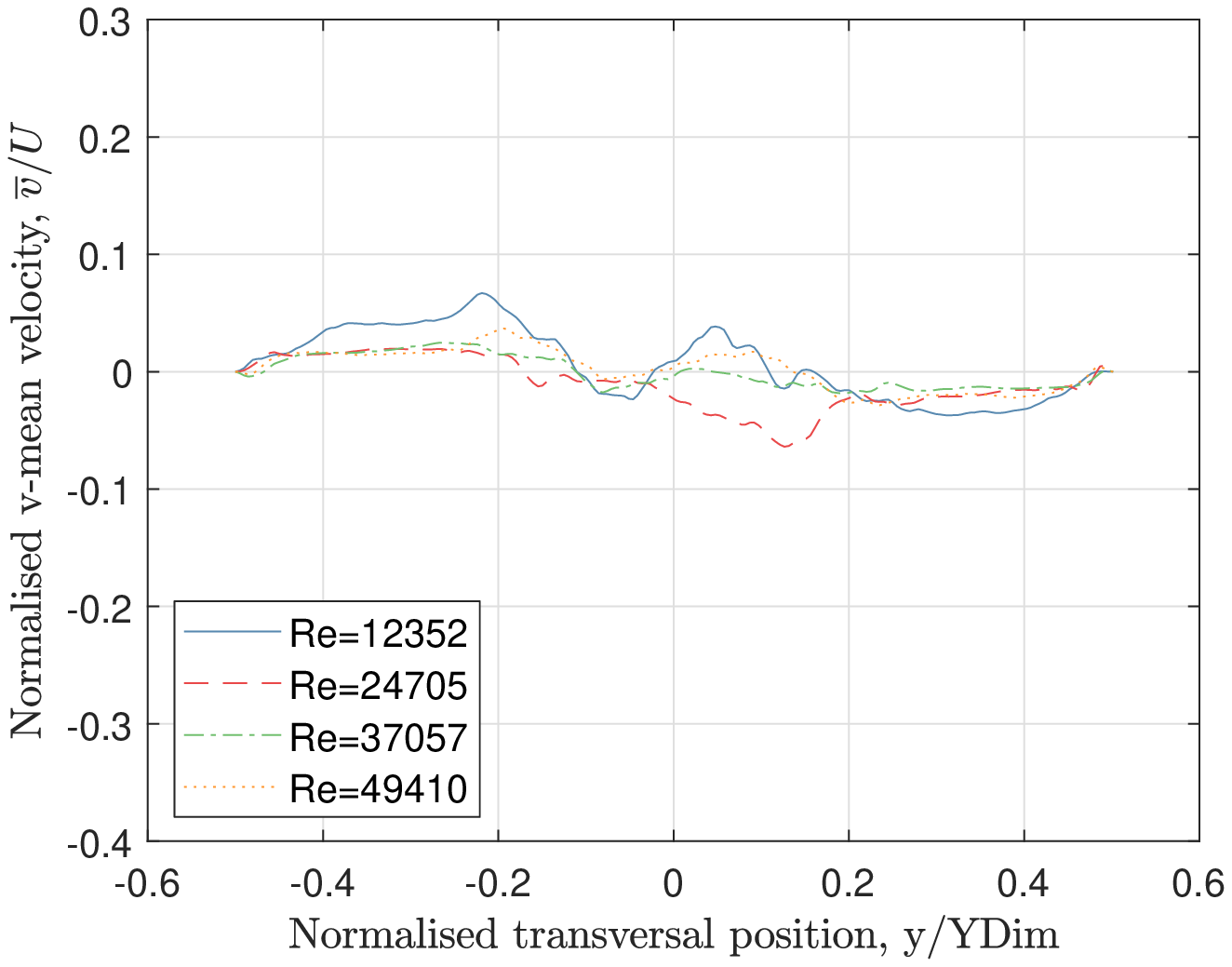}	
	\includegraphics[width=0.32\textwidth]{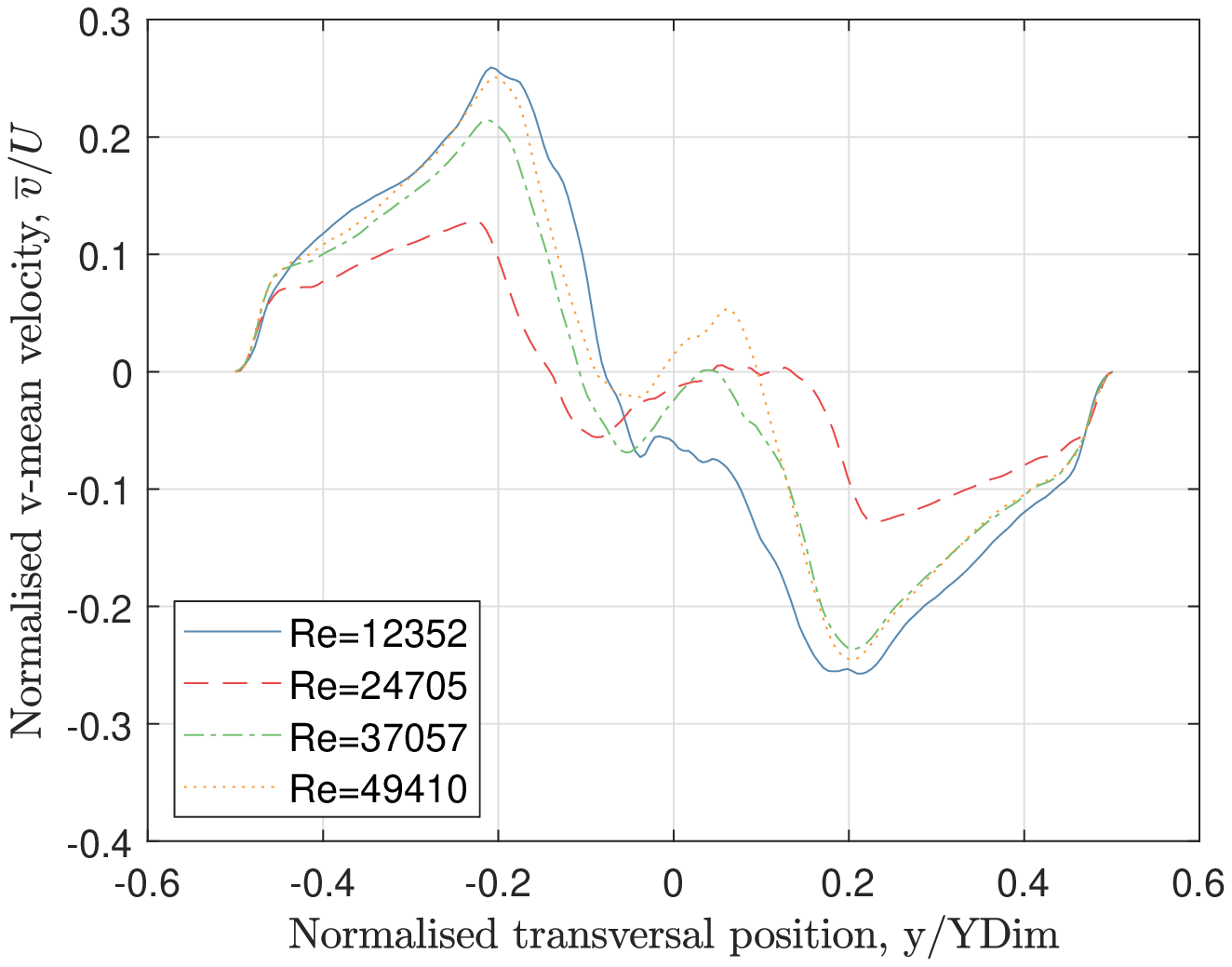}	
	\par\medskip	
	\includegraphics[width=0.32\textwidth]{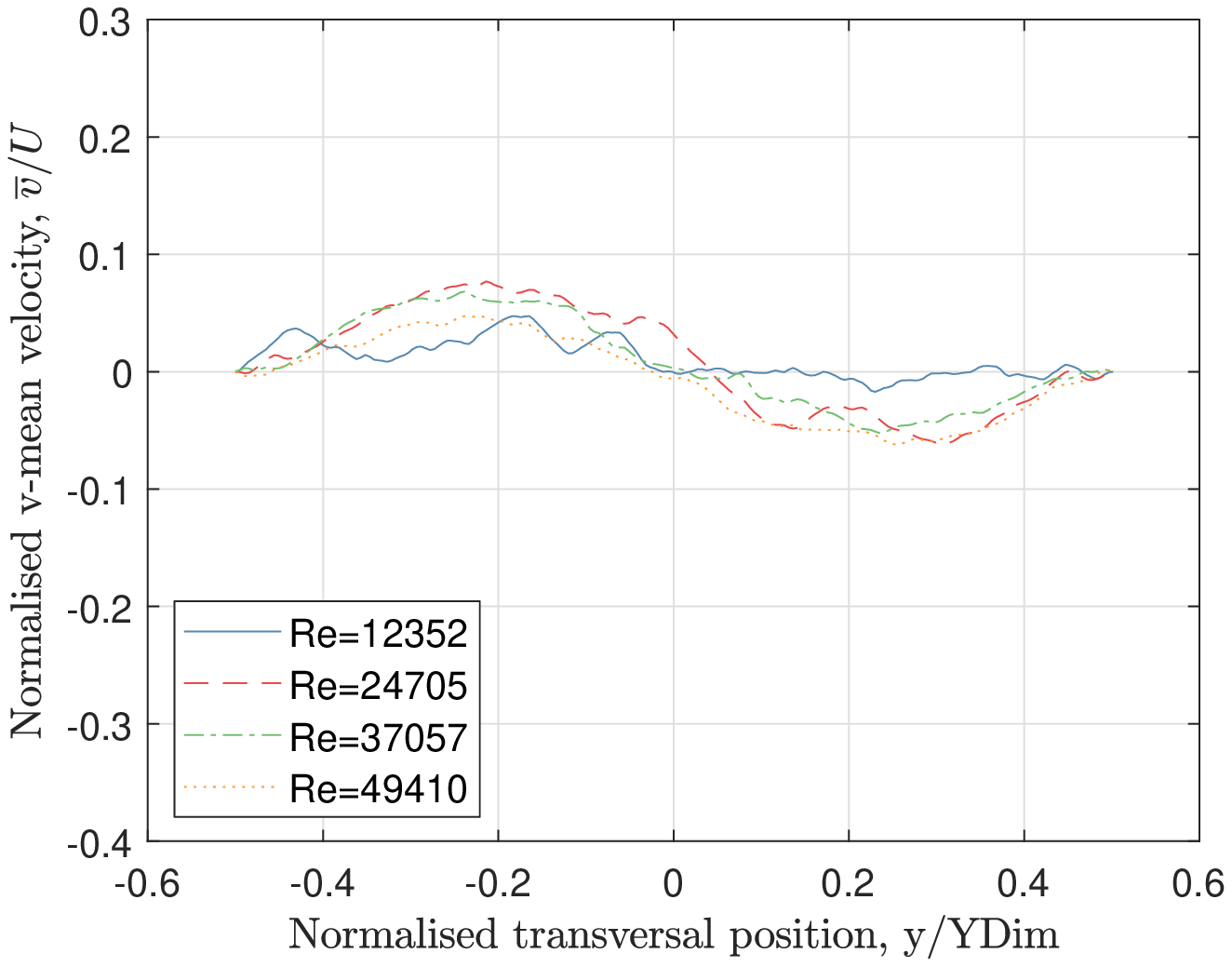}	
	\includegraphics[width=0.32\textwidth]{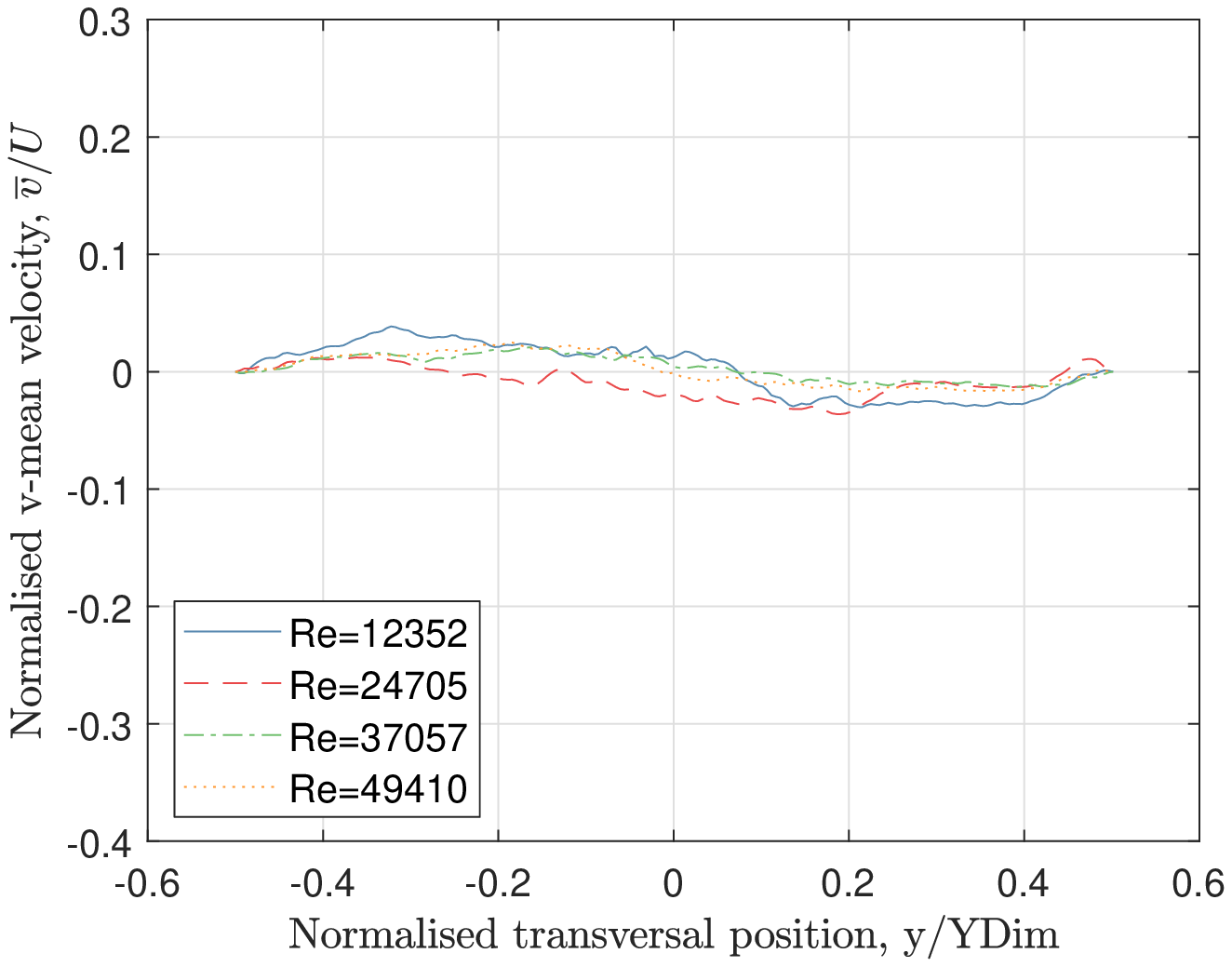}	
	\includegraphics[width=0.32\textwidth]{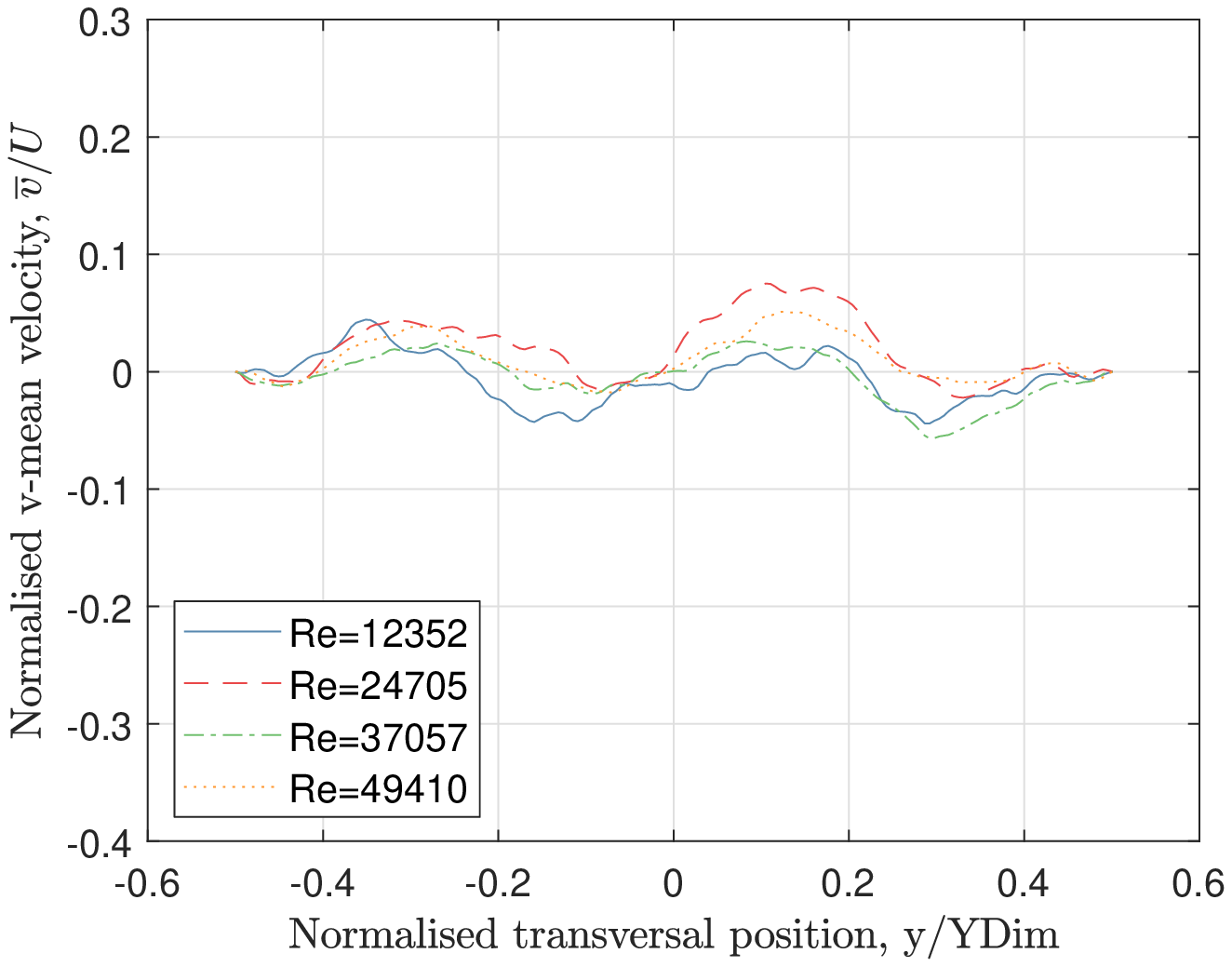}	
	\par\medskip	
	\caption{Transverse cross-stream velocity profiles (from top to bottom) 2D, 3D and 6D.}
	\label{MRT_Turbulent_MeanV_TransProfileImg}
\end{figure}
The transverse spanwise velocity ($v$) profiles for all three obstacles are shown in Fig.~\ref{MRT_Turbulent_MeanV_TransProfileImg} at different locations in the channel for the three obstacles (SS - Solid Square, PR - Porous Regular, PF - Porous Fractal). Each row represents a streamwise position after the obstacle. We only show the profiles at 2D, 3D and 6D from the obstacle as further down $v$ is not significant (Fig~\ref{MRT_Turbulent_MeanV_TransProfileImg}).

The cross-stream near-wake can be characterised quite simply as a clockwise rotating structure for the solid obstacle whereas quite interestingly for the regular porous obstacle there are two side by side clockwise structures with smaller amplitudes caused by the secondary flow in-between the smaller obstacle rows. In the case of the regular obstacle these structures are short lived and by 6D the flow is behaving as if the obstacle were a solid one. However, in the fractal case there is no well-defined structure in the near-wake but later at 3D we obsverve the clockwise rotating structure seen for the solid obstacle. 


\subsection{Strouhal Numbers}

The Strouhal number is defined as 
\begin{equation}
St = \frac{Df}{U_{\infty} }
\end{equation}
where $f$ is the velocity oscillation frequency measured locally using fast Fourier transforms. 
It is measured at 40\% flow depth from the channel floor.
\begin{figure}[ht]
	\centering			
	\includegraphics[width=0.45\textwidth]{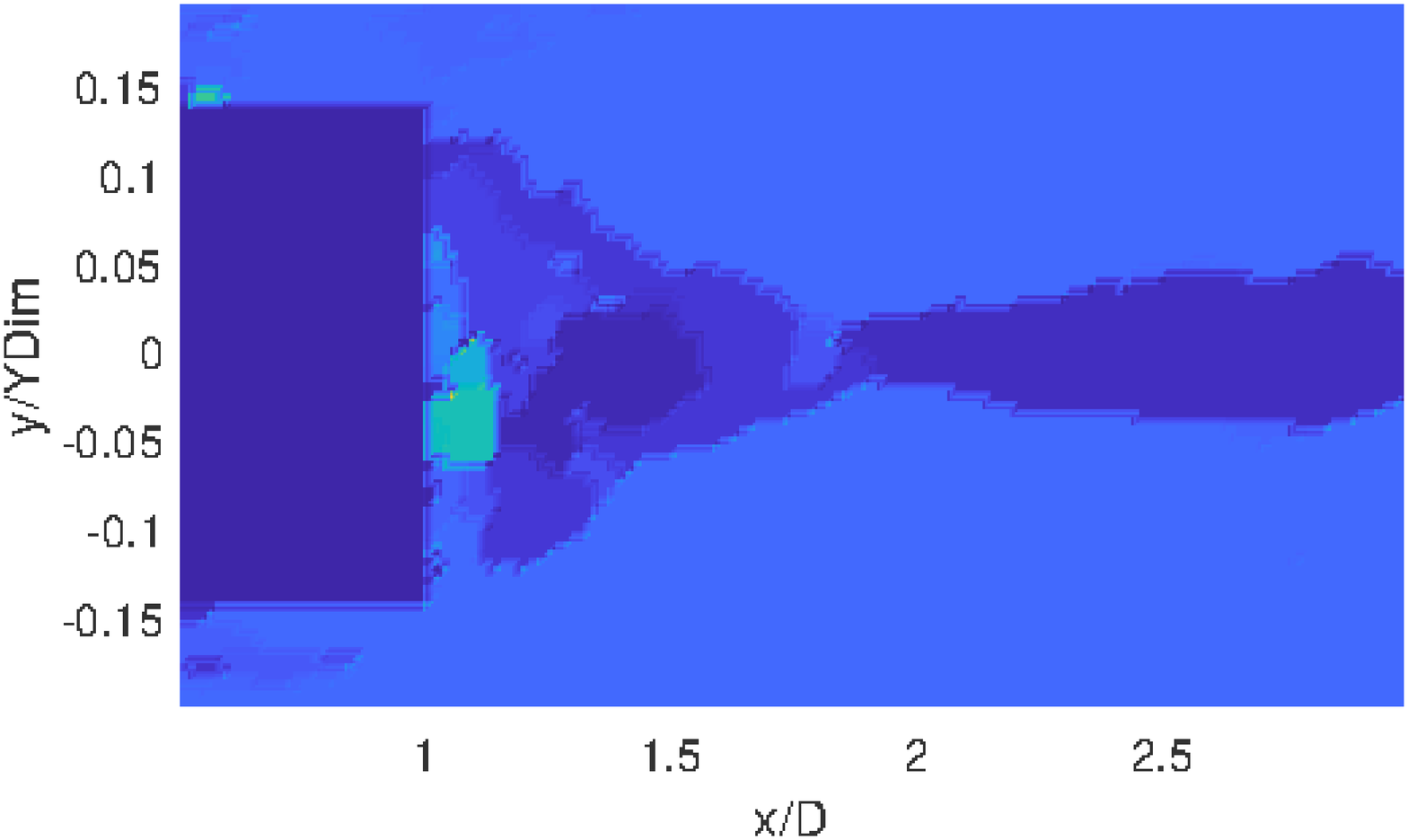}	
	\includegraphics[width=0.45\textwidth]{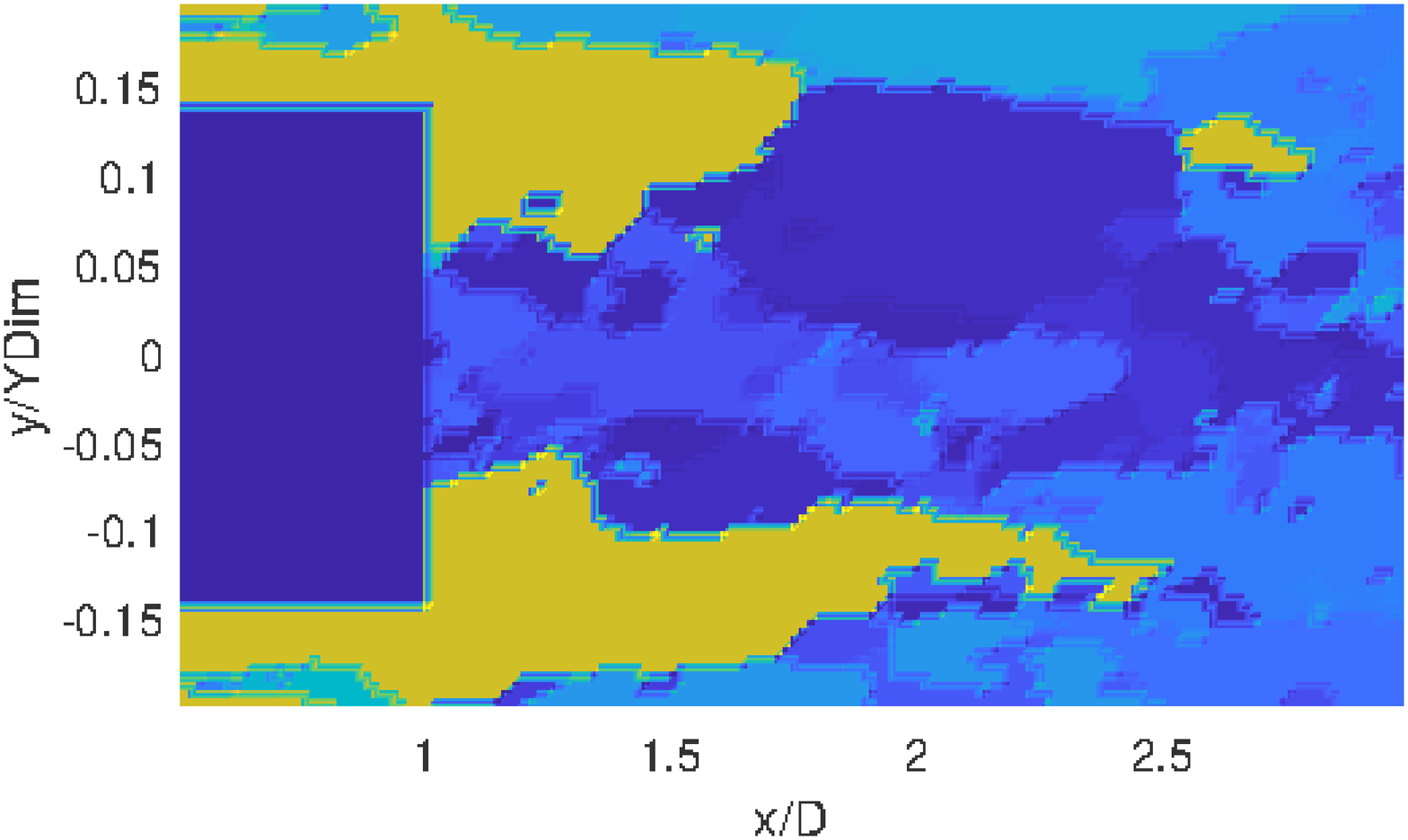}
	\par\medskip
	\includegraphics[width=0.45\textwidth]{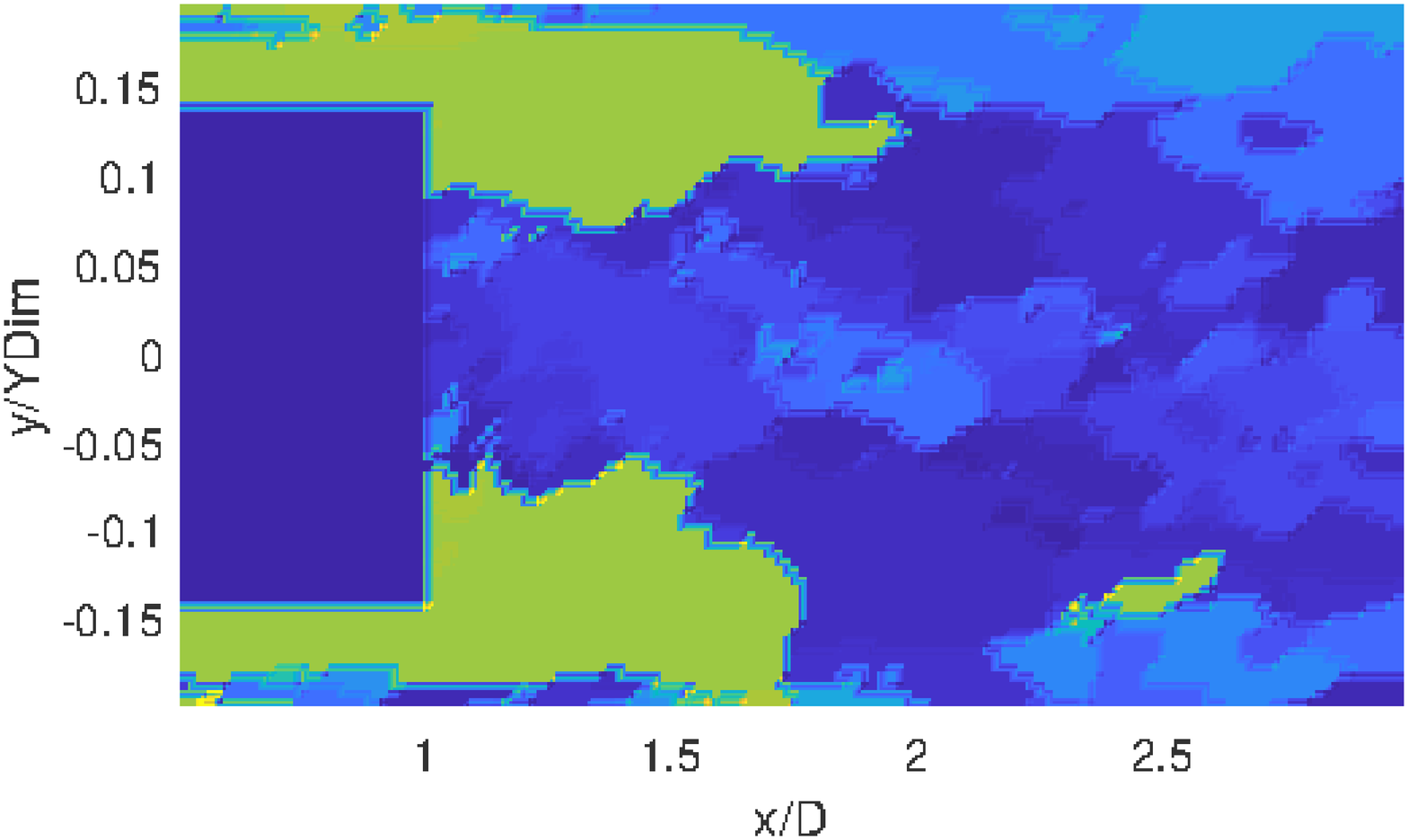}
	\includegraphics[width=0.45\textwidth]{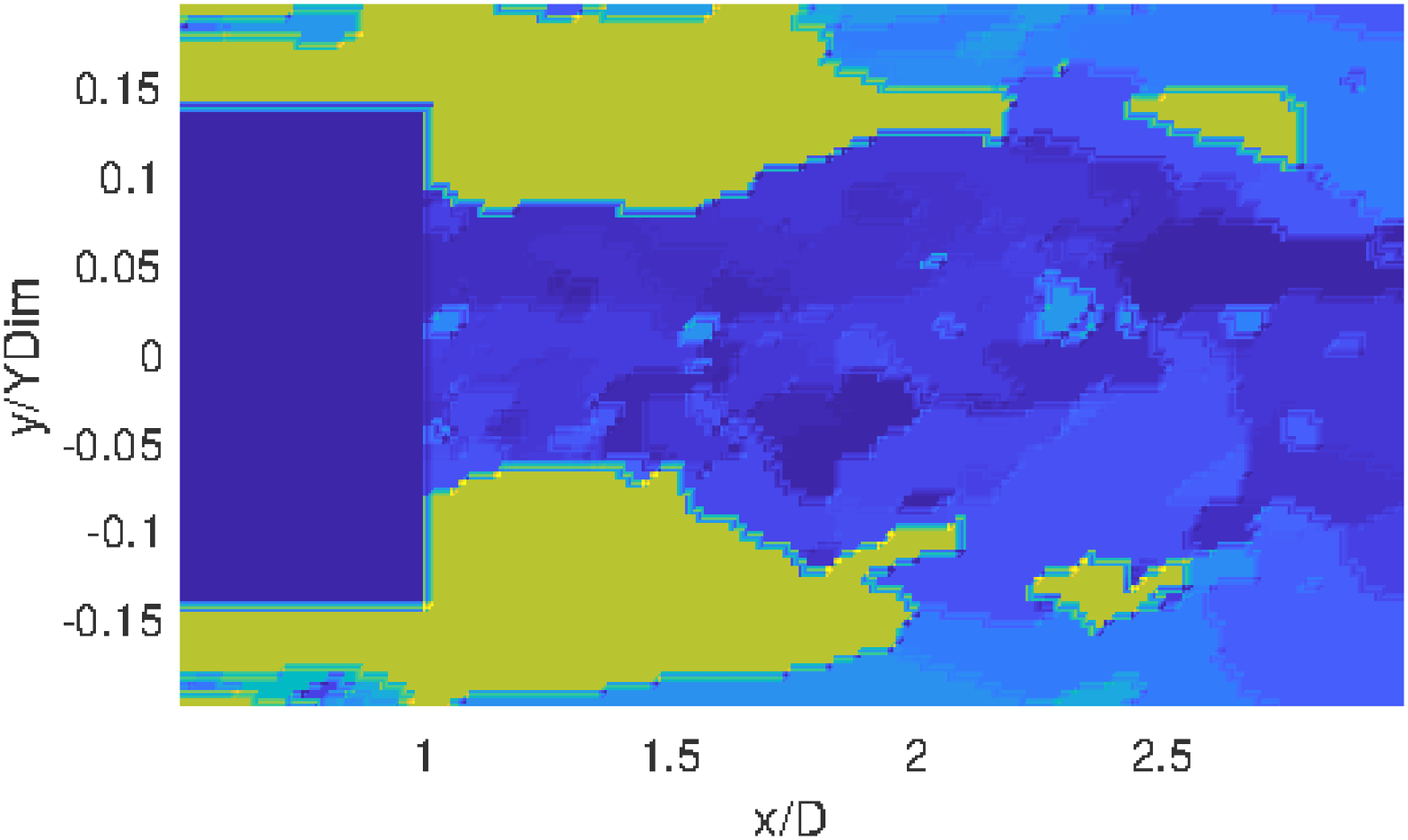}
	\par\medskip			
	\includegraphics[width=0.45\textwidth]{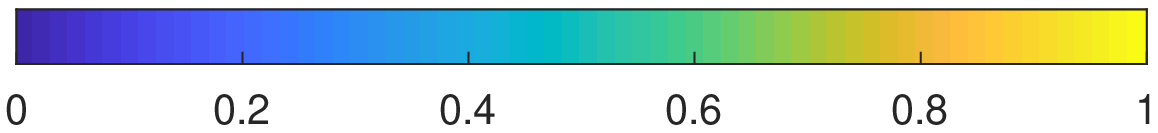}	
\caption{Strouhal number, $St$, maps of the Z-normal plane at 40\% flow depth from the channel floor. Solid square obstacle at (Top Left) $Re_D=12352$  (Top Right) $Re_D=24705$ (Bottom left) $Re_D=37057$ (Bottom right) $Re_D=49410$. }
	\label{SC_Strouhal_MapImg}
\end{figure} 

\begin{figure}[h]
	\centering
	\includegraphics[width=0.45\textwidth]{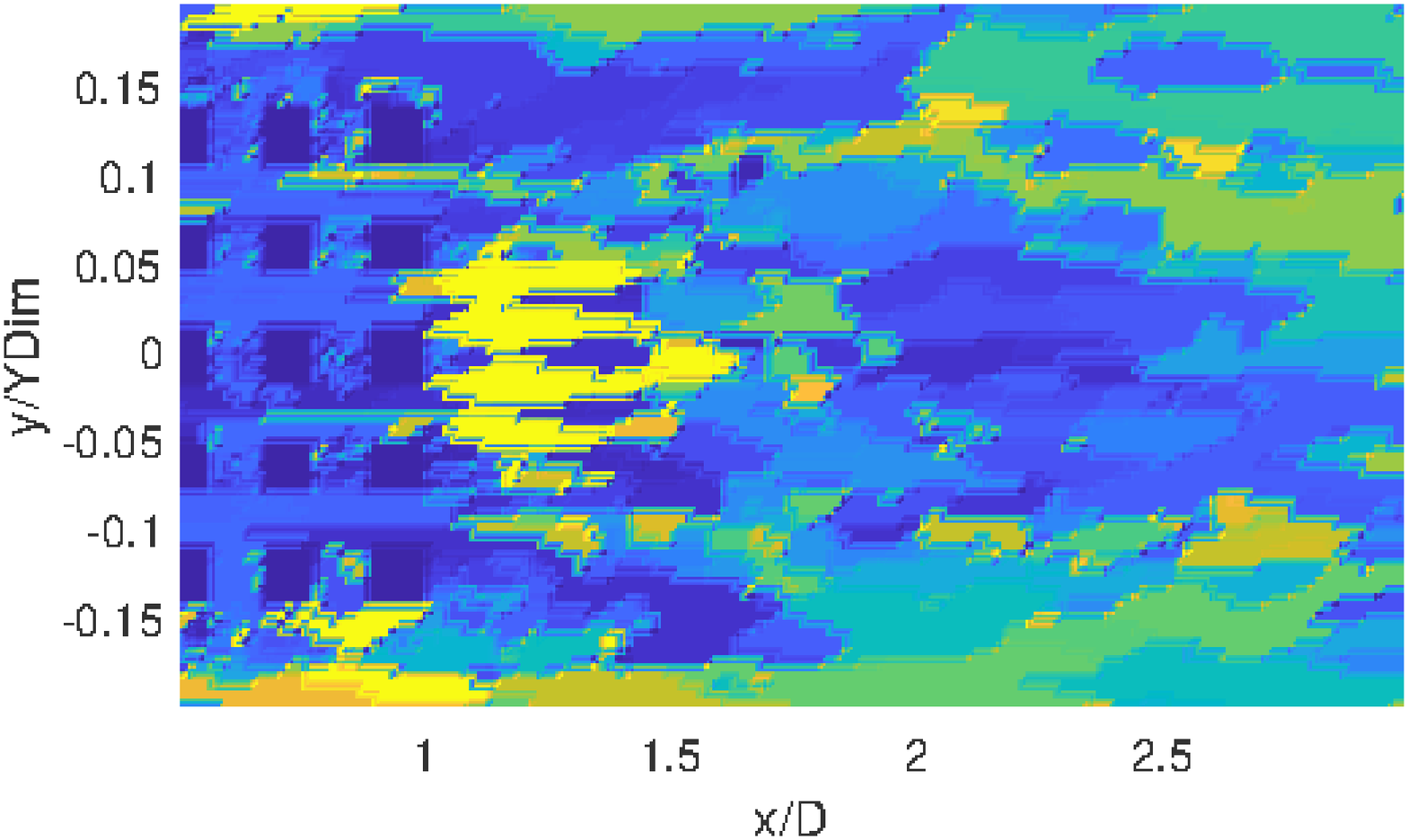}	
	\includegraphics[width=0.45\textwidth]{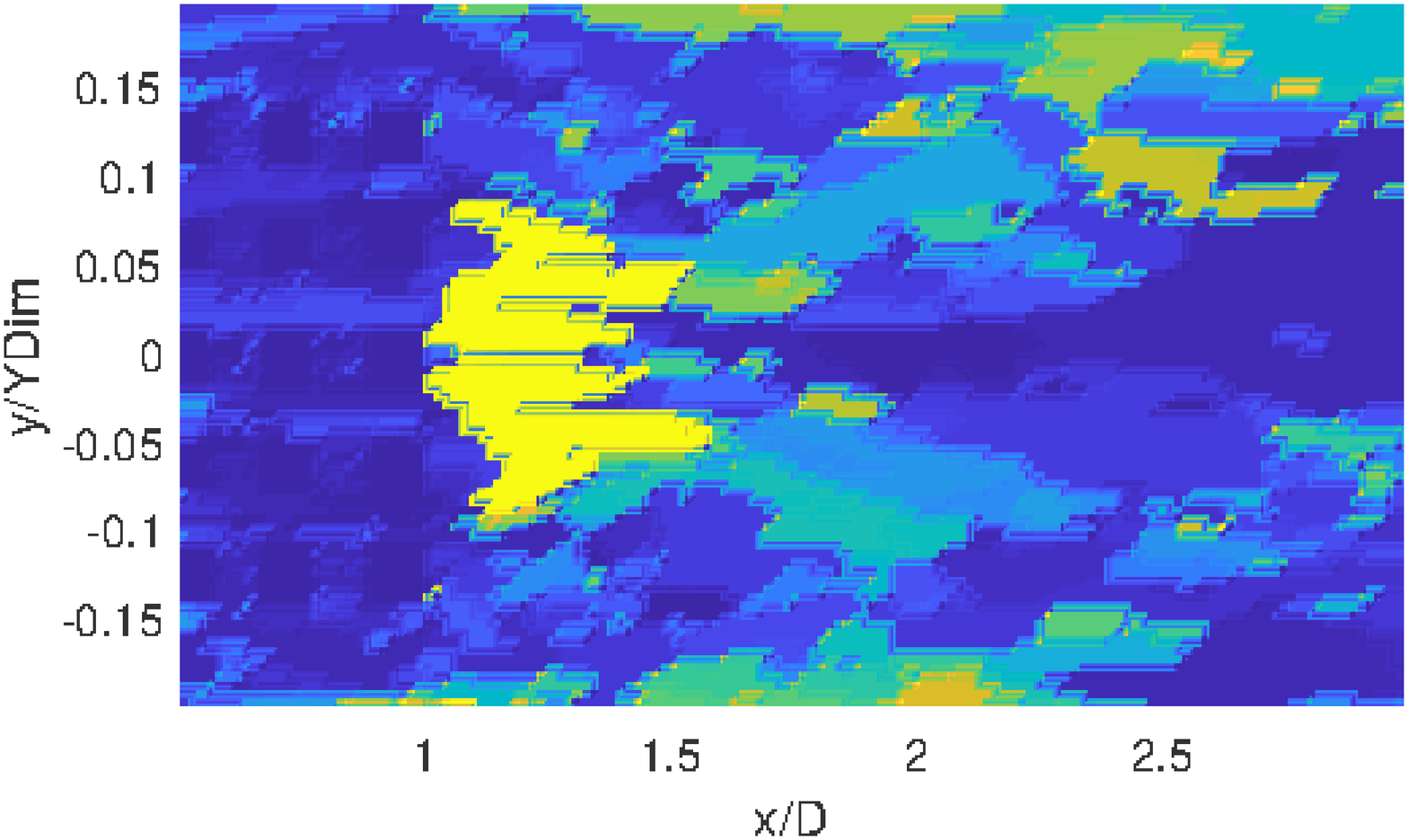}
	\par\medskip
	\includegraphics[width=0.45\textwidth]{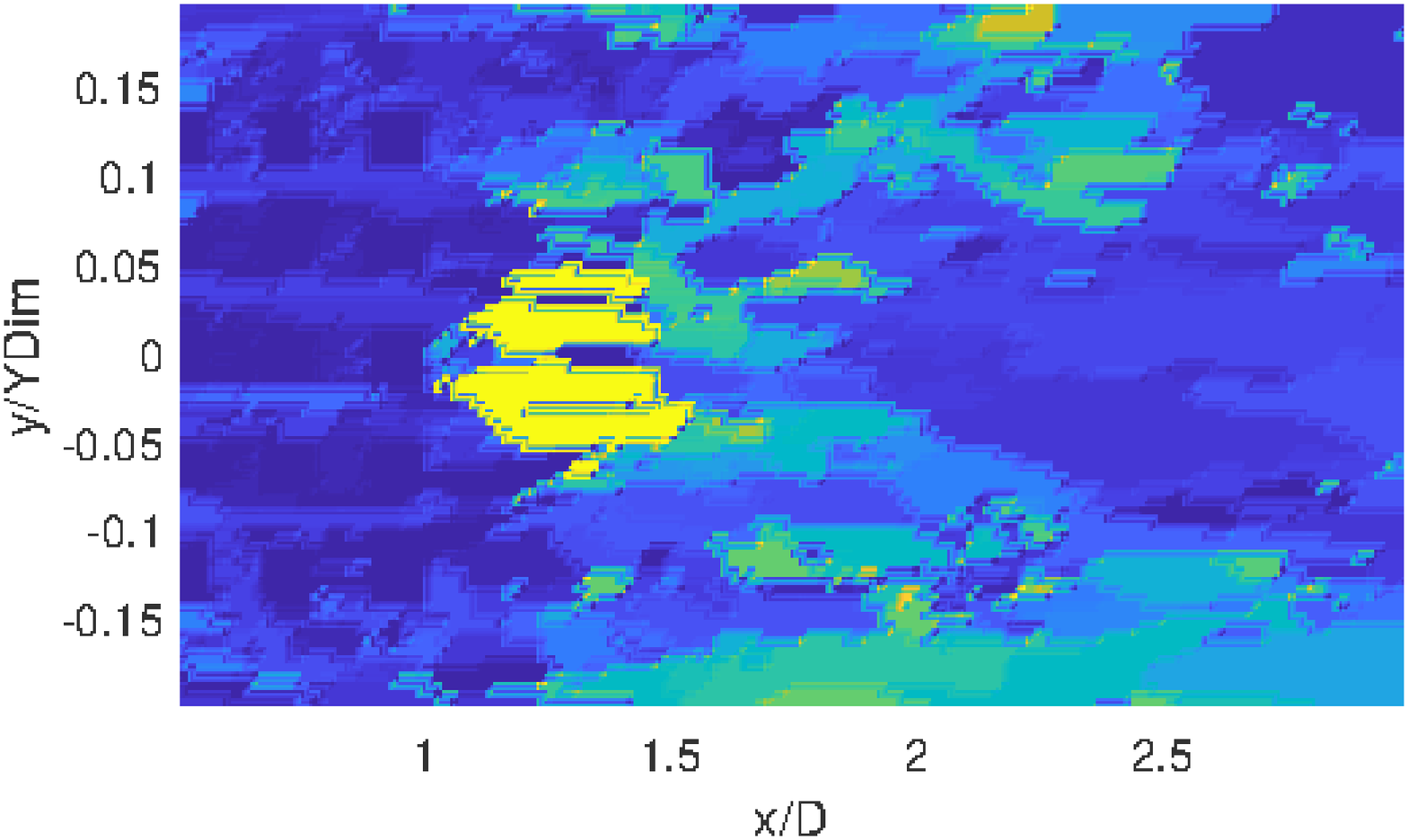}
	\includegraphics[width=0.45\textwidth]{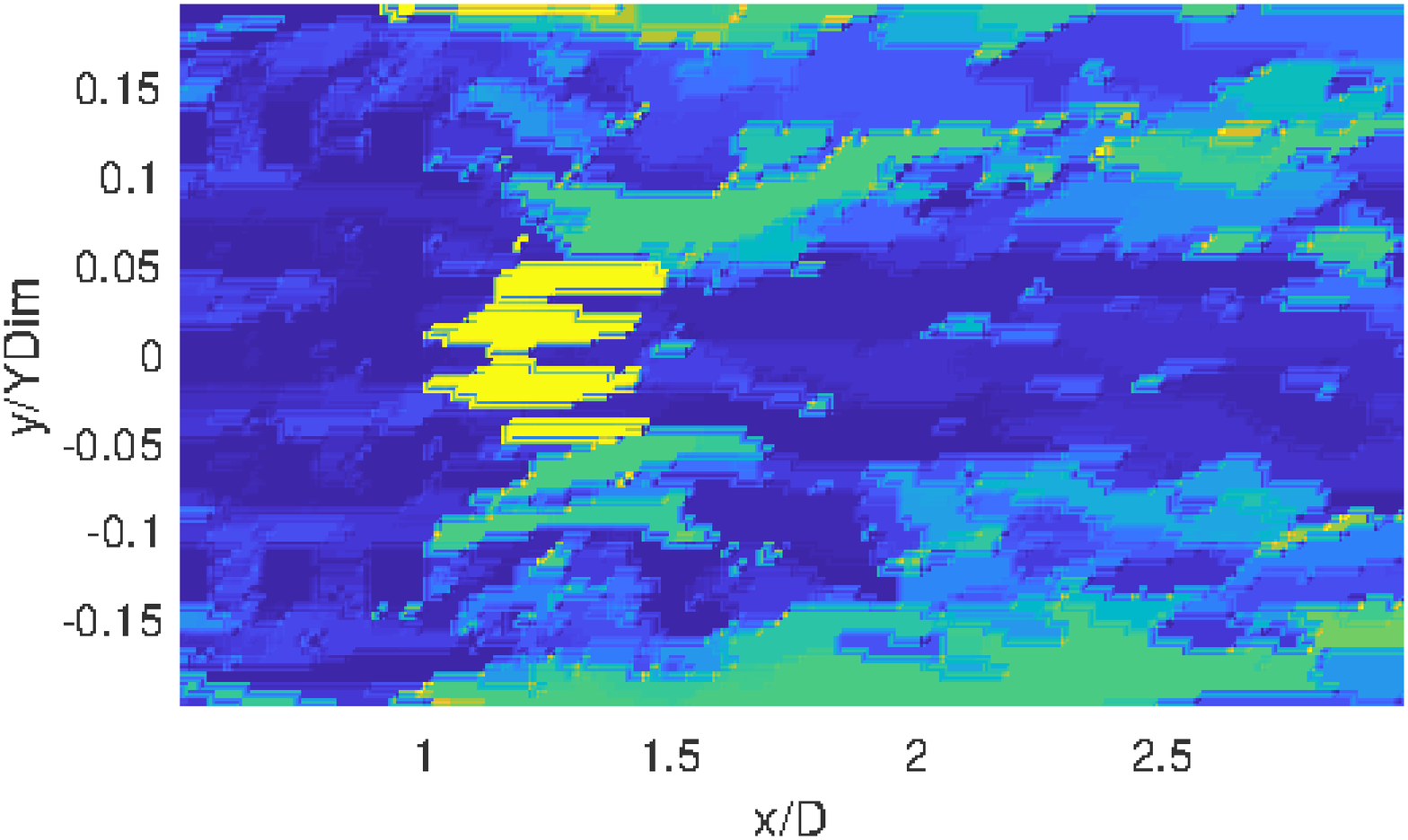}
	\par\medskip			
	\includegraphics[width=0.45\textwidth]{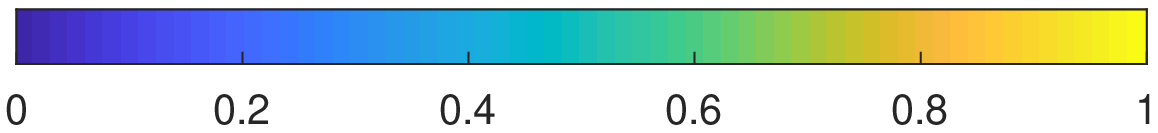}	
	\caption{Strouhal number, $St$, maps of the Z-normal plane at 40\% flow depth from the channel floor. Porous regular obstacle at (Top Left) $Re_D=12352$  (Top Right) $Re_D=24705$ (Bottom left) $Re_D=37057$ (Bottom right) $Re_D=49410$. }
	\label{RG_Strouhal_MapImg}
\end{figure}

\begin{figure}[h]
	\centering
	\includegraphics[width=0.45\textwidth]{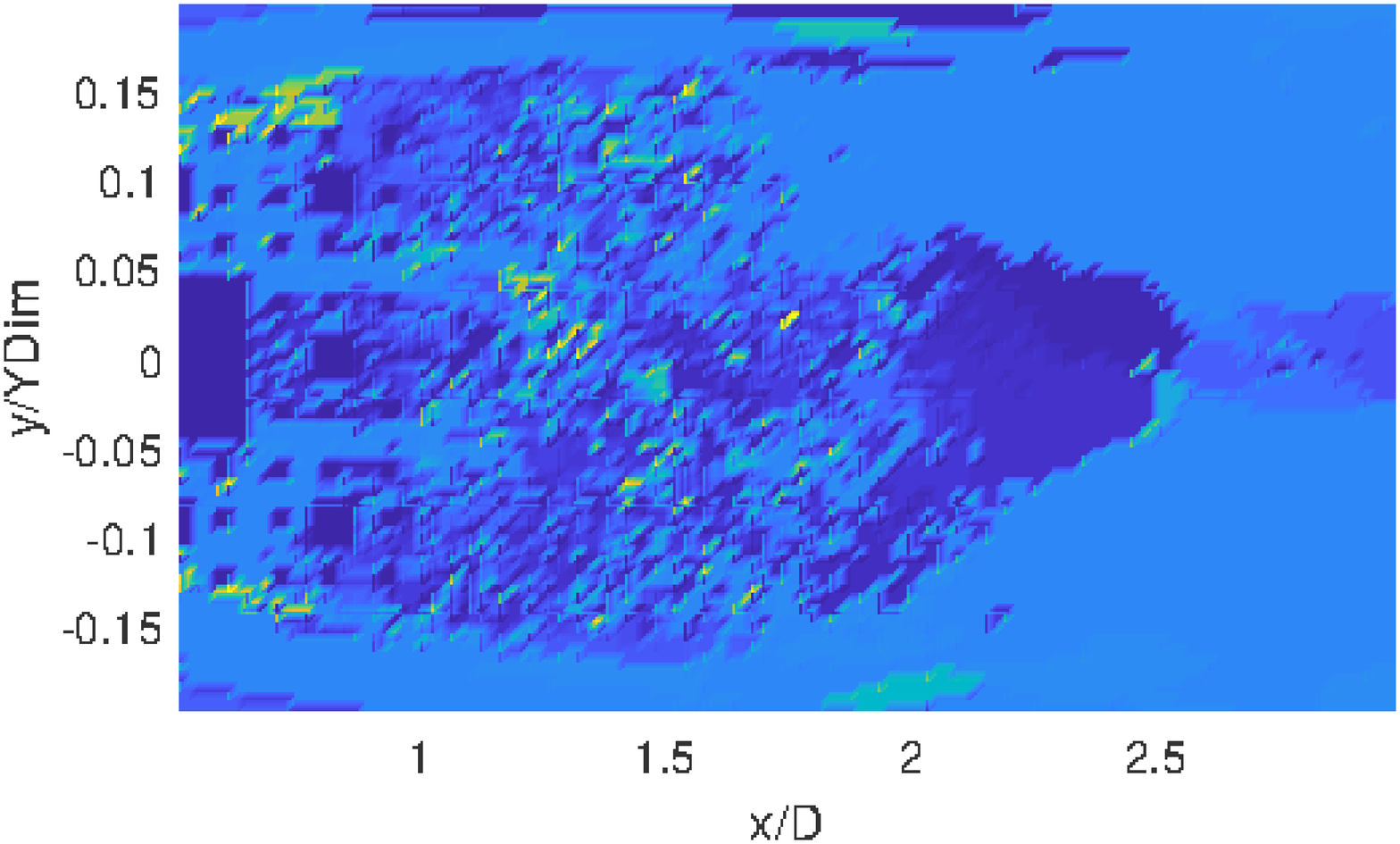}	
	\includegraphics[width=0.45\textwidth]{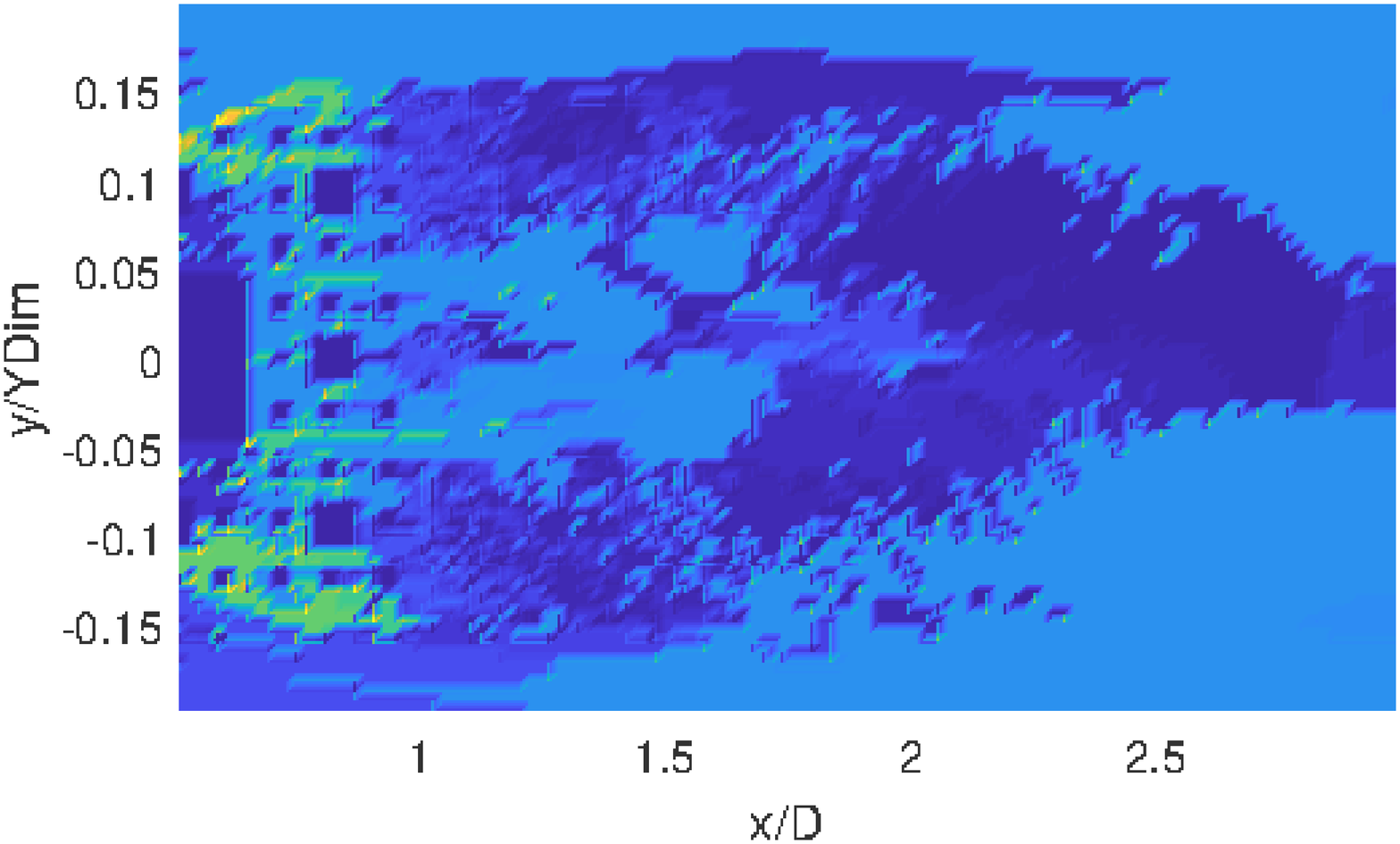}
	\par\medskip
	\includegraphics[width=0.45\textwidth]{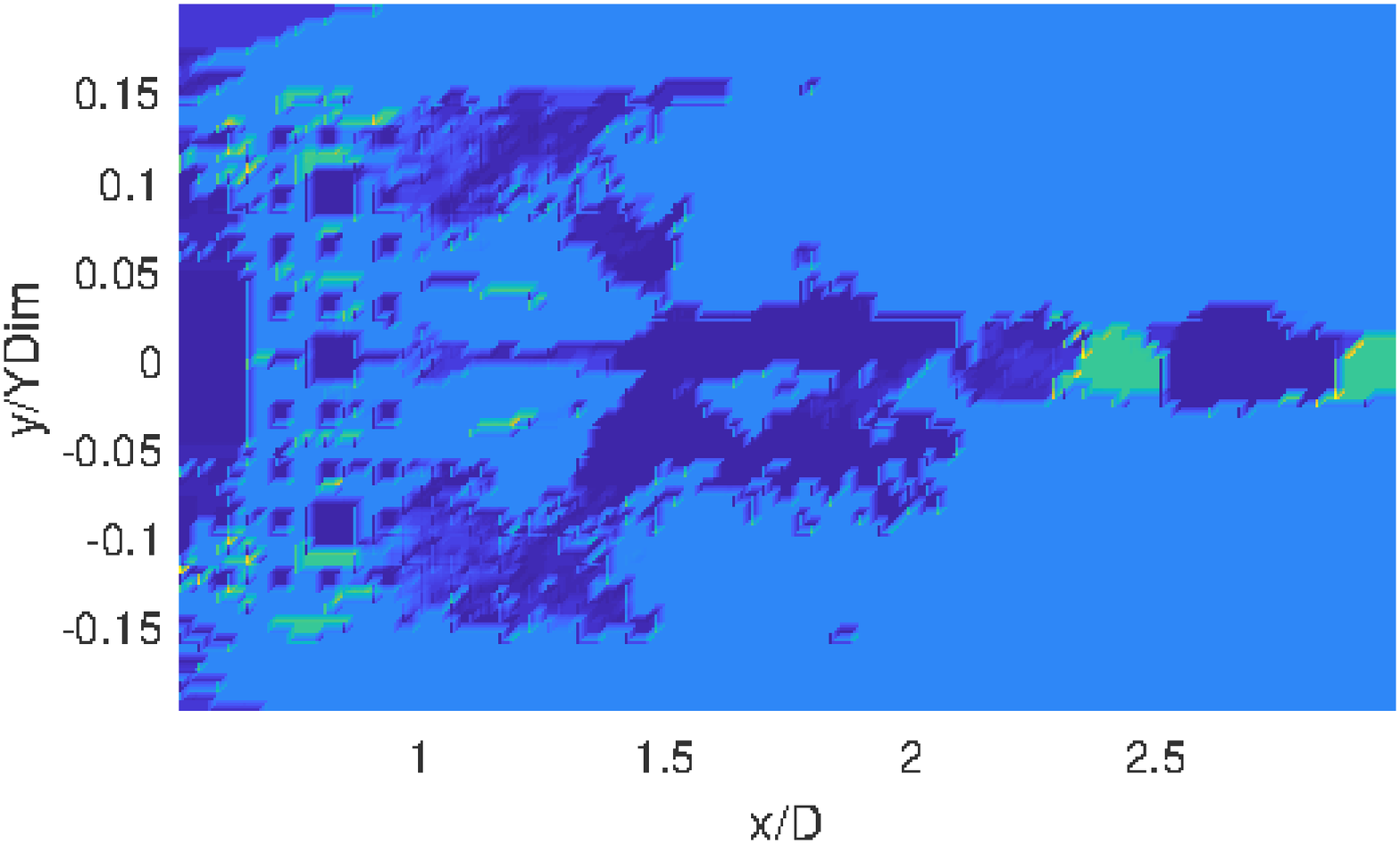}
	\includegraphics[width=0.45\textwidth]{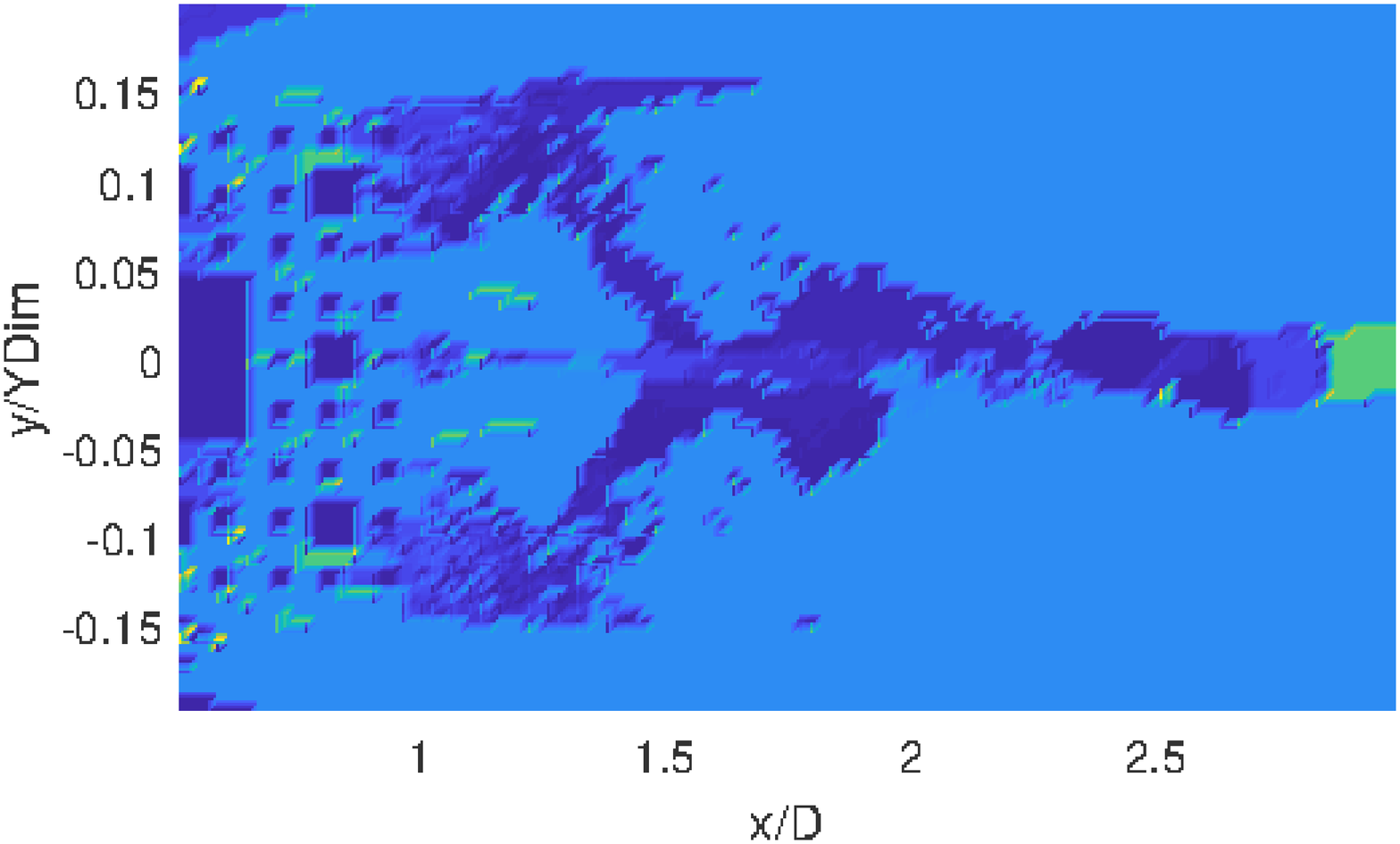}
	\par\medskip			
	\includegraphics[width=0.45\textwidth]{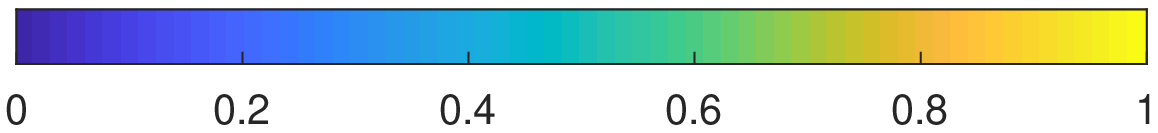}	
	\caption{Strouhal number, $St$, maps of the Z-normal plane at 40\% flow depth from the channel floor. Porous fractal obstacle at (Top Left) $Re_D=12352$  (Top Right) $Re_D=24705$ (Bottom left) $Re_D=37057$ (Bottom right) $Re_D=49410$. }
	\label{FC_Strouhal_MapImg}
\end{figure}

Maps of $St$ are shown for the solid square, porous regular, porous fractal objects in Figures~\ref{SC_Strouhal_MapImg}, \ref{RG_Strouhal_MapImg} and \ref{FC_Strouhal_MapImg} rspectively. The domain is cropped to show the area immediately downstream of the obstacle.
\\[2ex]
For the Strouhal numbers of the square obstacle, we can  look at  \citep{bosch1998simulation} for expected values, in which the Strouhal number is expected to range between 0.125-0.145. (See also \cite{Durao-et-al-1988} for a case at $Re = 14000$.) The Strouhal number maps presented in Fig.~\ref{SC_Strouhal_MapImg} agree with this in the area where the near wake ends. 
The lower Reynolds case exhibits a different behaviour whereas as commented earlier the larger Reynolds number cases tend toward a universal behaviour with similar pattern: two zones where $0.6 < St < 1$ corresponding to the vortices generated on each side of the square and a quieter zone in-between forming the wake.
\\[2ex]
For the porous regular obstacle (Fig.~\ref{RG_Strouhal_MapImg}), since there is no global wake region for the regular obstacle as there is little to impede the flow from going straight through, there is no clearly defined near wake. However, in the immediate area after the obstacle there is a region which has a relatively higher frequency compared to the surrounding region. This Strouhal number is roughly nine times ($St \simeq 1)$ the value for solid square obstacle. Considering that the regular obstacle comprises square cylinders which are nine times smaller than the solid obstacle it yields that this group of obstacles is not behaving as a group, instead the vortex shedding is being dominated by the individual cylinders. 
This region of high frequency becomes universal as the Reynolds number increases. 
\\[2ex]
The case of the fractal obstacle (Fig.~\ref{FC_Strouhal_MapImg}) is very different from that of the  porous regular obstacle though they share the same porosity. 
In particular there is no more this region of high frequency just after the obstacle and the group of obstacles is behaving as a group so that the group's wake property cannot be inferred from properties of each individual elements in the obstacle. 

In the case of the fractal obstacle, since there is a defined wake region, by looking at the edge of the near wake we again see similar scaling effect instead this time the vortex shedding is being dominated by the largest cylinder in the obstacle. 
Once again the map tends toward a universal behaviour as the Reynolds number increases.


\subsection{Wake Length}
\begin{figure}[h]
	\centering
	\includegraphics[width=0.5\textwidth]{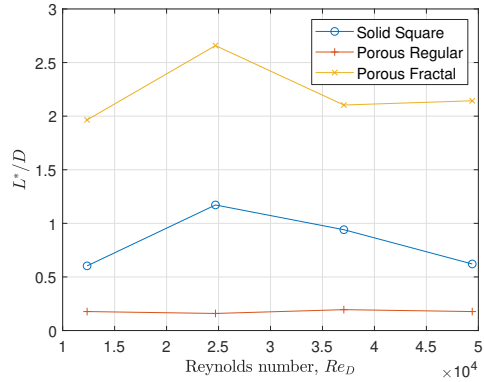}
	\caption{Length to minimum streamwise velocity, $L^*$, from the base of the obstacle in the streamwise direction for all three obstacles}
	\label{L*_AllObstaclesImg}
\end{figure}

Given that the regular obstacle does not form a wake, wake lengths can only truly be compared between the fractal and solid obstacle. Additionally, owing to the fractal delaying the formation of the recirculation zone two definitions could be adopted, one being the length from the base of the obstacle to the point where the mean streamwise velocity remains positive or the maximum length where the mean streamwise velocity is negative. Therefore, to adequately compare the three different obstacles a parameter common to all three is defined as follows: $L^*$ is the length from the base of the obstacle to the minimum mean streamwise velocity. Fig.~\ref{L*_AllObstaclesImg}, shows the variation in this length for all the tested obstacles and Reynolds numbers.
\\[2ex]
Overall, the flow is least impeded by the porous regular obstacle, given that along the centreline there are no sub-obstacles this is to be expected. However, the fractal obstacle having one third of the porosity of the solid obstacle has a greater impeding effect on the flow. In all cases both the solid and fractal cases show a similar pattern, the maximum length to the minima occurs at $Re_D=24705$. In the case of the solid obstacle it then steadily decreases with increasing flowrate, whilst for the fractal obstacle $L^*$ remains constant after $Re_D=37057$. This further demonstrates the earlier conclusion that although the near wake region is increased by the use of a fractal obstacle the recovery is much faster with increasing Reynolds numbers. 

\section{Conclusion}
\label{seconcl}

We simulated velocity fields from turbulent flows over four different obstacles:
Solid (SS), Porous Regular (PR) and Porous Fractal (FR), using LBM.
The flow was simulated for four Reynolds numbers and compared to ADV experimental results.

LBM predicts very well the mean streamwise velocity and TKE profiles for the porous regular obstacle.

It predicts well the profiles for the fractal obstacle except 
in the far wake for the mean streamwise velocity. 

Normalised profiles achieve a universal behaviour faster for the porous obstacles. For the square obstacle it is achieved for $Re > 24000$.

In the case of a fractal obstacle it is paramount to check that the smallest fractal scale is properly meshed (Fig.~\ref{FC_MeshImg}). The local capture of the smallest geometry is more important than its fine meshing.

As expected the wake observed after the obstacle depends greatly on the internal structure of the porous object. The oscillation observed for the solid square are annihilated in the case of the porous regular obstacle but only pushed downstream in the case of the porous fractal object.




\begin{thebibliography}{11}
\newcommand{\enquote}[1]{#1}
\providecommand{\natexlab}[1]{#1}
\providecommand{\url}[1]{\texttt{#1}}
\providecommand{\urlprefix}{URL }
\expandafter\ifx\csname urlstyle\endcsname\relax
  \providecommand{\doi}[1]{doi:\discretionary{}{}{}#1}\else
  \providecommand{\doi}{doi:\discretionary{}{}{}\begingroup
  \urlstyle{rm}\Url}\fi

\bibitem[{Bosch and Rodi(1998)}]{bosch1998simulation}
Bosch, G. and Rodi, W. [1998] \enquote{Simulation of vortex shedding past a
  square cylinder with different turbulence models,} \emph{International
  journal for numerical methods in fluids} \textbf{28}(4),  601--616.

\bibitem[{Coleman and Vassilicos(2008)}]{Coleman-Vassilicos-2008}
Coleman, S.~W. and Vassilicos, J.~C. [2008] \enquote{Transport properties of
  saturated and unsaturated porous fractal materials,} \emph{Physical Rev.
  Let.} \textbf{100}(035504).

\bibitem[{d'Humieres(1992)}]{d1992generalized}
d'Humieres, D. [1992] \enquote{Generalized lattice-boltzmann equations,}
  \emph{Rarefied gas dynamics} ,  450--458.

\bibitem[{d{\textquoteright}Humi{\`e}res(2002)}]{Humieres2002MRT}
d{\textquoteright}Humi{\`e}res, D. [2002]
  \enquote{Multiple{\textendash}relaxation{\textendash}time lattice boltzmann
  models in three dimensions,} \emph{Philosophical Transactions of the Royal
  Society of London A: Mathematical, Physical and Engineering Sciences}
  \textbf{360}(1792),  437--451, \doi{10.1098/rsta.2001.0955}.

\bibitem[{Dur{\~a}o \emph{et~al.}(1988)Dur{\~a}o, Heitor and
  Pereira}]{Durao-et-al-1988}
Dur{\~a}o, D., Heitor, M. and Pereira, J. [1988] \enquote{Measurements of
  turbulent and periodic flows around a square cylinder,} \emph{Experiments in
  Fluids} \textbf{6},  298--304.
\bibitem[{Higham and Brevis(2018)}]{Higham2001Modification}
Higham, J. and Brevis, W. [2018] \enquote{Modification of the modal
  characteristics of a square cylinder wake obstructed by a multi-scale array
  of obstacles,} \emph{Experimental Thermal and Fluid Science} \textbf{90},
  212--219.

\bibitem[{Higham \emph{et~al.}(2016)Higham, Brevis and
  Keylock}]{Higham-et-al-2016}
Higham, J.~E., Brevis, W. and Keylock, C.~J. [2016] \enquote{A rapid
  non-iterative proper orthogonal decomposition based outlier detection and
  correction for piv data,} \emph{Meas. Sci. Technol.} \textbf{27},  125303.

\bibitem[{Laizet and Vassilicos(2012)}]{Laizet-Vassilicos-2012}
Laizet, S. and Vassilicos, J.~C. [2012] \enquote{Fractal space-scale unfolding
  mechanism for energy-efficient turbulent mixing,} \emph{Phys Rev E}
  \textbf{86},  046302.

\bibitem[{Latt(2009)}]{latt2009palabos}
Latt, J. [2009] \enquote{Palabos, parallel lattice boltzmann solver,} .

\bibitem[{Nicolleau \emph{et~al.}(2011)Nicolleau, Salim and
  Nowakowski}]{Nicolleau-et-al-JOT-2011}
Nicolleau, F., Salim, S. and Nowakowski, A. [2011] \enquote{Experimental study
  of a turbulent pipe flow through a fractal plate,} \emph{Journal of
  Turbulence} \textbf{12}(44),  1--20, doi: 10.1080/14685248.2011.637046.

\bibitem[{Smagorinsky(1963)}]{smagorinsky1963general}
Smagorinsky, J. [1963] \enquote{General circulation experiments with the
  primitive equations: I. the basic experiment*,} \emph{Monthly weather review}
  \textbf{91}(3),  99--164.

\bibitem[{Wang \emph{et~al.}(2016)Wang, Nicolleau and Qin}]{Wei-et-al-2016-FDR}
Wang, W., Nicolleau, F. C. G.~A. and Qin, N. [2016] \enquote{Comparison of
  turbulent flow through hexagram and hexagon orifices in circular pipes using
  large-eddy simulation,} \emph{Fluid Dynamics Research} \textbf{48}(2),
  021408, \doi{10.1088/0169-5983/48/2/021408},

\end{thebibliography}

\end{document}